\documentclass{article}
\usepackage{slashed,verbatim}
\usepackage[numbers,sort&compress]{natbib}
\usepackage{amsmath}
 
\renewcommand \thesubsection{\Roman{section}.\Alph{subsection}}

\newcommand{\Rmnum}[1]{\expandafter\@slowromancap\romannumeral #1@}
\makeatother
\usepackage{amssymb}
\usepackage{float}
\usepackage{appendix}
\usepackage{graphicx}
\usepackage{subcaption}
\usepackage{verbatim}
\usepackage{dcolumn}
\usepackage{bm}
\usepackage[colorlinks=true,linktocpage=true,
linkcolor=blue,citecolor=blue]{hyperref}
\usepackage[a4paper]{geometry}
\newcommand{\be}{\begin{equation}}
	\newcommand{\ee}{\end{equation}}

\begin{document}
	\large
	\title{\bf{Dynamics of heavy flavour in a weakly magnetized hot QCD medium}}
	\author{Debarshi Dey\footnote{ddey@ph.iitr.ac.in, debs.mvm@gmail.com}~~and~~Binoy Krishna
		Patra\footnote{binoy@ph.iitr.ac.in}\vspace{0.1in}\\
		Department of Physics,\\
		Indian Institute of Technology Roorkee, Roorkee 247667, India}
	\maketitle
	
	\begin{abstract}
		We obtain the spatial and momentum diffusion coefficients ($D_s$ and $\kappa$), and 
		the collisional energy loss ($dE/dx$) of a heavy quark (HQ) traversing through a thermal medium of quarks and gluons in a weak magnetic field ($B$),  for the two cases of the HQ moving either parallel or perpendicular to $\bm{B}$. For that purpose, we consider Coulomb scatterings ($t$-channel) of the HQ with the light quarks, obtained from the imaginary part of the HQ self-energy via the cutting rules. Both the normalised (by $T^3$) 
		 $\kappa$, as well as $dE/dx$, for charm quarks are larger than that for bottom quarks due to the larger mass of the latter.  Also, the effect of $B$ is more feeble on the bottom quark, compared to the charm quark. Comparatively, the magnitudes of both $\kappa$ and $dE/dx$ are significantly smaller for the case of $\bm{v}\perp\bm{B}$. For both the cases, our results show that the momentum transfer between the HQ and the medium takes place preferentially along the direction of HQ velocity, thus leading to a significant increase in the momentum diffusion anisotropy, compared to $B=0$.
		We also calculate the (scaled) spatial diffusion coefficient, which we find to be independent of the heavy flavor mass and is almost unaffected by changes in $B$. 
	\end{abstract}
	
	\section{INTRODUCTION}	
	Heavy-ion collisions at experimental facilities such as the Relativistic Heavy Ion Collider (RHIC) and the Large Hadron Collider (LHC) have presented strong evidence of formation of a deconfined thermal QCD medium, called the Quark Gluon Plasma (QGP)\cite{Arsene:NPA757'2005,Adams:NPA757'2005}. When the heavy nuclei collide non-centrally, the spatial asymmetry of the initial overlap zone is carried over to the momenta of emitted particles, and can be seen experimentally in the final hadron $p_T$ spectra\cite{QGP3,Shuryak:NPA750'2005,Huovinen:ARNPS56'2006,Hirano:JPG36'2009}. This asymmetric expansion of the QGP fireball is referred to as elliptic flow. A remarkable property of this medium is the very small value of the ratio of shear viscosity to entropy density $\eta/s$, making QGP one of the most perfect fluids known\cite{Kovtun:PRL95'2005}. A microscopic explanation of these interesting transport properties is still a subject of intense investigation. To that end, heavy quarks (Charm, Bottom) are considered to be excellent probes of the QGP medium\cite{Frawley:PR462'2008}.  The large mass of the heavy quark $M_Q$, compared to the temperature of the medium ($M_Q\gg T$) means that heavy quarks are formed at very early stages of heavy-ion collisions, even before the formation of the thermal medium\cite{Levai:PRC51'1995}. Typical heavy quark (HQ) formation time is $\sim 1/2M_Q$\cite{Das:PLB768'2019} ($\sim$0.08 fm/c for Charm, $\sim$0.03 fm/c for Bottom). Further, compared to the light quarks, the HQ thermal relaxation times are larger, parameterically by a factor $M_Q/T\sim 5$-15. The light quark and gluon thermalization time $\tau_{q,g}$ is $\sim$ 0.6 fm, as indicated by hydrodynamic modelling of the RHIC data\cite{Heinz:ACP739'2004}. This implies a HQ thermalisation time of 3-9 fm/c, which is of the order of (for Charm) or larger (for Bottom) than the estimated QGP lifetime of $\sim$ 5 fm/c\cite{He:arxiv'2022}. Because of being created so early, and not equilibrating fully with the medium, the HQs are witnesses to the full history of space-time evolution of the medium and also retain a ``memory" of their interactions with the medium. This makes heavy quarks ideal probes of the QGP medium. Extensive discussions on HQ phenomenology can be found in Refs.\cite{Rapp:QGP4'2010,He:arxiv'2022}. 

	Apart from causing an anisotropic expansion of the created matter, non-central heavy ion collisions also lead to creation of large magnetic fields\cite{Tuchin:AdvHEP'2013}. The decay rate of the magnetic field depends strongly on the electrical conductivity of the medium which is exposed to the field\cite{Tuchin:PRC82'2010,Tuchin:PRC83'2011,Marty:PRC88'2013,Ding:PRD83'2011,Gupta:PLB597'2004,Amato:PRL111'2013,Aarts:PRL99'2007,Puglisi:PRD90'2014,Greif:PRD90'2014,Hattori:PRD96'2017}. 
	Assuming  a large background magnetic field, several phenomena have been studied 
	such as chiral magnetic effect (CME)\cite{Kharzeev:NPA803'2008},chiral magnetic wave\cite{Kharzeev:PRD83'2011,Newman:JHEP01'2006}, charge dependent elliptic flow\cite{Burnier:PRL107'2011,Gorbar:PRD83'2011}, magnetic catalysis (MC)\cite{Alexandre:PRD63'2001,Gusynin:PRD56'1997,Lee:PRD55'1997}, inverse magnetic catalysis (IMC)\cite{Bali:JHEP02'2012,Farias:PRC90'2014,Farias:EPJA53'2017,Mueller:PRD91'2015,Ayala:PRD90'2014,Ayala:PRD91'2015,Ayala:PLB759'2016}, etc.   For small conductivities however, the magnetic field decay would be very fast. This has led to studies of QGP transport properties in a weak background magnetic field in the recent past\cite{Gupta:PLB597'2004,Aarts:JHEP02'2015,Ding:PRD94'2016,Rath:PRD100'2019,Khan:PRD104'2021,Li:PRD97'2018,Panday:arxiv,Kurian:PRD103'2021,Khan:PRD107'2023,Dey:PRD104'2021,Hasan:PRD102'2020}. For the case of HQs, their production times are small($\sim 1/2M_Q$). This is similar to the timescale of generation of (strong) magnetic field. However, in the plasma frame, the heavy quarks are formed at a time $t_f (=\gamma \tau_f)$ which could be of the order of 1-2 $fm$, depending on the momenta of the produced HQ. By that time, the strength of the magnetic field may become weak.
	Further, although HQ transport has been studied in the presence of a strong background magnetic field, using both imaginary and real time formalism\cite{Fukushima:PRD93'2016,Singh:JHEP68'2020,Bandyopadhyay:PRD105'2022,Jamal:arxiv}, the literature using weak background magnetic field is rather scant. This motivates us to investigate the dynamics of individual heavy flavour in the QGP medium, in the limit of a weak background magnetic field. 
	In the context of heavy quarks, the diffusion of HQ in a thermal medium has been studied using perturbation theory\cite{Thoma:NPB351'1991,Braaten:PRD44'1991,Moore:PRC71'2005,Monteno:JPG38'2011,He:PLB735'2014,Beraudo:EPJC75'2015,Das:PLB747'2015,Sadofyev:PRD93'2016,Akamatsu:PRC92'2015,Andronic:EPJC76'2016,Kurian:PRD100'2019,Singh:arxiv}, lattice QCD\cite{Banerjee:PRD85'2012,Bouttefeux:JHEP150'2020,Banerjee:arxiv,Caron-Huot:JHEP04'2009}, in a Polyakov-loop plasma\cite{Singh:PRD100'2019}.  HQ dynamics in the presence of external EM fields has been studied in \cite{Das:PLB768'2019, Chatterjee:PLB798'2019}, where, in the former, the authors show that the directed flow ($v_1$) of HQs is a good probe of the magnetic field generated in non-central HICs. In the latter study, the combined effect of the initial tilt of the QGP fireball and large EM fields on the HQ $v_1$ is studied. Effects of the initial pre-equilibrium glasma phase on HQ observables has been explored in \cite{EPJA54'2018,Liu:PRD105'2021}. HQ drag and diffusion in strongly coupled plasmas has been studied in \cite{Casalderrey-Solana:PRD74'2006,Rajagopal:JHEP10'2015}. Next-to-leading order (NLO) calculation of the HQ diffusion has also been carried out\cite{Caron-Huot:PRL100'2008}. Effect of momentum anisotropy on the dynamics of HQ has also been studied recently\cite{Kumar:PRC105'2022,Prakash:arxiv}. Recently, a non-perturbative study of HQ diffusion in strong magnetic fields has been carried out\cite{Mazumder:arxiv}
	
	The HQ mass is the hardest scale in the problem, which is true even if the magnetic field is strong. The scale hierarchy considered in this problem is $M_Q\gg T\gg eB/T$.  We calculate the energy loss $dE/dx$, and momentum diffusion coefficients $\kappa$ of the HQ moving with a finite velocity in the QGP by evaluating the scattering rate of the HQ with the light thermal quarks and gluons. In particular, we consider two cases of the HQ velocity: one in which the HQ velocity is parallel to $\bm{B}$, and other, in which the HQ velocity lies in a plane perpendicular to $\bm{B}$.
	Cutting rules allow for determination of this scattering rate from the imaginary part of the HQ self energy\cite{Weldon:PRD28'2007}. This method was employed to study HQ dynamics for the first time in\cite{Braaten:PRD44'1991}. In what follows, the HQ self energy is evaluated using an effective gluon propagator, which, in turn, is calculated in the presence of a weak magnetic field, up to second order in $qB$. HTL perturbation theory is made use of throughout the calculations. Owing to its large mass, the problem of HQ immersed in a thermal bath of light particles is amenable to a non-relativistic treatment, in general, and a diffusion treatment, in particular, as will be justified in the next section. We go beyond the  static heavy quarks and evaluate the aforementioned quantities for finite HQ momentum.
	
	The paper is organized as follows: In section II the description of heavy quarks in a thermal medium is presented, wherein, the case of zero magnetic field is discussed first, followed by the case of a finite magnetic field. Then in section III, the calculation of the scattering rate $\Gamma$ in the presence of a weak magnetic field is outlined. In section IV, the evaluation of energy loss $dE/dx$ and momentum diffusion coefficients $\kappa$ are presented,  first for the case $\bm{v}\parallel \bm{B}$, followed by the case $\bm{v}\perp \bm{B}$. In section V, the results obtained are discussed. In section VI, phenomenological applications of the calculations are described. Finally, we conclude in section VII.
	
	\section{DESCRIPTION OF HEAVY QUARKS IN A THERMAL MEDIUM: STATIC CASE AND BEYOND}
	\subsection{Kinematics in the absence of magnetic field}
	We consider a heavy quark of mass $M_Q$ propagating through a plasma of light quarks and gluons. The HQ thermal momentum  $p$ $\sim$ $\sqrt{M_QT}\gg T$ translates to a thermal velocity $v\sim \sqrt{T/M_Q}\ll 1$. Even if one considers hard scatterings of the HQ with the light medium particles (characterised by a momentum transfer of $\mathcal{O}(T)$), it takes a large number of collisions ($\sim M_Q/T$) to change the HQ momentum by $\mathcal{O}(1)$, since $p\gg T$. This implies that the momentum changes accumulate over time from uncorrelated ``kicks", and the HQ momentum therefore evolves according to Langevin dynamics:
	\begin{equation} \label{eq1}
		\frac{dp_i}{dt}=\xi_i(t)-\eta_Dp_i\,, \qquad \langle \xi_i(t)\xi_j(t')\rangle=\kappa\,\delta_{ij}\delta(t-t'),
	\end{equation}
	where, $(i,j)=(x,y,z)$. These are the macroscopic Langevin equations with $\eta_D$ being the momentum drag coefficient and $\kappa$ the momentum diffusion coefficient. The random forces $\xi(t)$ representing the uncorrelated momentum kicks are assumed to be white noises. 
	The solution of Eq.\eqref{eq1} under the assumption $\eta_D^{-1}\ll t$ is given as
	\begin{equation}\label{eq2}
		p_i(t)=\int_{-\infty}^{t}dt'\,e^{\eta_D(t'-t)}\xi_i(t').
	\end{equation}
	$\kappa$ can be determined by calculating the mean squared momentum transfer per unit time from the underlying microscopic theory:
	\begin{equation}\label{eq3}
		\langle p^2\rangle=\int dt_1dt_2e^{\eta_D(t_1+t_2)}\langle\xi_i(t_1)\xi_j(t_2)\rangle=\frac{3\kappa}{2\eta_D}.
	\end{equation}
	Equivalently, $\kappa$ can be defined as
	\begin{equation}\label{eq3a}
		3\kappa(\bm{p})=\lim _{\Delta t \rightarrow 0} \frac{\left\langle(\Delta p)^2 \right\rangle}{\Delta t},
	\end{equation}
	where, $\Delta p=p(t+\Delta t)-p(t)$. This leads to the following equations of motion for the heavy quark.
	\begin{align}
		\frac{d}{dt}\langle p\rangle &\equiv -\eta_D(p)p\\
		\frac{1}{3}\frac{d}{dt} \left\langle(\Delta p)^2 \right\rangle&\equiv \kappa(p)
	\end{align}
The drag coefficient or the relaxation rate $\eta_D$ is related to $\kappa$ via the fluctuation-dissipation relation, 
\begin{equation}\label{drag}
	\eta_D=\frac{\kappa}{2M_QT},
\end{equation}\label{eq4}
which follows from general thermodynamical arguments. Apart from $\eta_D$ and $\kappa$, we also have the spatial diffusion coefficient $D_s$ and the heavy quark energy loss $dE/dx$. The problem of HQ motion and its subsequent diffusion in a thermal medium can be characterised fully by these 4 quantities, which are related to each other. In particular, 
\begin{equation}\label{eq5}
	D_s=\frac{T}{M_Q\,\eta_D}=2T^2/\kappa,
\end{equation} 
as derived in\cite{Reif:book}. 
	In a thermal medium of light quarks and gluons the random momentum kicks originate from the scattering processes $qH\rightarrow qH$ and $gH\rightarrow gH$ ($q\rightarrow$ quark, $g\rightarrow$ gluon). The former occurs only via $t$ channel Coulomb scattering. The latter, effectively also occurs via the same mechanism since its Compton amplitude is suppressed by $v^2\sim T/M_Q$, in the rest frame of the plasma. This is especially true for the bottom quark (M=4.18 GeV), compared to the charm quark (M=1.28 GeV). 
We assume that the dominant mechanism for the HQ energy loss is coulomb scattering of the HQ with the light medium particles and ignore radiative energy loss (gluon bremsstrahlung), which is suppressed by an additional power in the strong coupling $\alpha_s$, as explained in\cite{Moore:PRC71'2005}. The central quantity from which all the above mentioned dynamical quantities can be obtained is the scattering rate $\Gamma$, whose computation will be outlined in the next section. The energy loss and the momentum diffusion coefficient are given as
	\begin{align}
		\frac{d E}{d x} & =\frac{1}{v} \int d^3 q \frac{d \Gamma(q)}{d^3 q} q_0 \label{dedx}\\[0.3em]
		3\kappa & = \int d^3 q \frac{d \Gamma(q)}{d^3 q} q^2.\label{kappa}
	\end{align}
	$\frac{d \Gamma(q)}{d^3 q}$ is the  differential probability per unit time for the heavy quark momentum to change by $\bm{q}$. It can also be interpreted as the scattering rate of heavy quark via one-gluon exchange with thermal partons per unit volume of momentum transfer $\bm{q}$. $v$ is the heavy quark velocity. $q_0$ and $q$ are respectively the energy and 3-momentum (magnitude) of the exchanged gluon. The factor of 3 comes from assuming isotropicity of the momentum diffusion coefficient, which is valid if the heavy quark under consideration is assumed to be static and the background magnetic field is weak.

	Beyond the case of static heavy quarks, the motion of the HQ along a particular direction leads to the generalized Langevin equations:
	\begin{equation} \label{Langevin}
		\frac{dp_i}{dt}=\xi_i(t)-\eta_Dp_i\,, \qquad \langle \xi_i(t)\xi_j(t')\rangle=\kappa_{ij}(\bm{p})\,\delta(t-t'),
	\end{equation}
	where, $\kappa_{ij}(\bm{p})=\kappa_L(p)\hat{p_i}\hat{p_j}+\kappa_T(p)(\delta_{ij}-\hat{p_i}\hat{p_j})$. $\kappa_L$ and $\kappa_T$ are the longitudinal and transverse momentum diffusion coefficients, respectively. The direction of HQ velocity
	 defines an anisotropy direction and the momentum diffusion coefficient breaks into longitudinal and transverse components as $3\kappa\rightarrow\kappa_L+2\kappa_T$. The factor 2 reflects the fact that there are two equivalent transverse directions.
	\begin{align}
		\kappa_L&=\int d^3 q \frac{d \Gamma(q)}{d^3 q} q_L^2\\[0.2em]
		\kappa_T&=\frac{1}{2}\int d^3 q \frac{d \Gamma(q)}{d^3 q} q_T^2
	\end{align}
	The HQ momentum can be diffused via collisions, in the direction of HQ momentum and also transverse to it, of which, $\kappa_L$ and $\kappa_T$ respectively are quantitative measures. $\kappa_L$ and $\kappa_T$ will separately satisfy Einstein relations as follows:
	\begin{equation}
		(\eta_D)_L=\frac{\kappa_L}{2M_QT}\,,\qquad (\eta_D)_T=\frac{\kappa_T}{2M_QT}.
	\end{equation}

\subsection{Kinematics with a finite magnetic field}
 The presence of a magnetic field introduces an additional scale in the system. In this work, the strength of the magnetic field is considered to be weak, that is, the scale hierarchy satisfies $M_Q\gg T\gg eB/T$. The direction of the external magnetic field causes an anisotropy in the momentum diffusion coefficients, similar to what was discussed earlier. Considering the case of static HQ, and taking the direction of the external magnetic field to be along $\hat{z}$ direction,  one obtains the following Langevin equations corresponding to directions parallel and perpendicular to the magnetic field\cite{Fukushima:PRD93'2016}.
\begin{align}
	\frac{dp_{z}}{dt}&=-(\eta_D)_{\parallel}p_z+\xi_z\,,\quad \langle \xi_z(t)\xi_z(t')\rangle=\kappa_{\parallel}(\bm{p})\,\delta(t-t')\\
	\frac{d\bm{p}_{\perp}}{dt}&=-(\eta_D)_{\perp}\bm{p}_{\perp}+\bm{\xi}_{\perp}\,,\quad \langle \bm{\xi}_{\perp}(t)\bm{\xi}_{\perp}(t')\rangle=\bm{\kappa}_{\perp}(\bm{p})\,\delta(t-t'),
\end{align} where, $\eta_z$ and $\bm{\eta_{\perp}}$ are the components of random forces parallel and perpendicular to $\bm{B}$., and the momentum diffusion coefficients are obtained as
\begin{align}
	\kappa_{\parallel}&=\int d^3 q \frac{d \Gamma(E)}{d^3 q} q_{\parallel}^2\\[0.2em]
	\bm{\kappa_{\perp}}&=\frac{1}{2}\int d^3 q \frac{d \Gamma(E)}{d^3 q} \bm{q_{\perp}}^2
\end{align}
The longitudinal and transverse drag and diffusion coefficients separately satisfy fluctuation-dissipation relations, as earlier
\begin{equation}
	(\eta_D)_{\parallel}=\frac{\kappa_{\parallel}}{2M_QT}\,,\qquad (\eta_D)_{\perp}=\frac{\bm{\kappa_{\perp}}}{2M_QT}.
\end{equation}
Now, if the HQ also has a velocity along a certain direction, then one needs to consider the interplay between the directions of HQ velocity and magnetic field.

\subsubsection{Case 1: $v\parallel B$}
In the first case, we consider the HQ velocity to be along the direction of magnetic field, so that, effectively, we still have a single preferred direction in space ($\hat{z}$). Then, one obtains the longitudinal and transverse momentum diffusion coefficients as
\begin{align}
\kappa_{L}&=\int d^3 q \frac{d \Gamma(E,v)}{d^3 q} q_{z}^2\\[0.2em]
\bm{\kappa_{\perp}}&=\frac{1}{2}\int d^3 q \frac{d \Gamma(E,v)}{d^3 q} \bm{q_{\perp}}^2
\end{align}

\subsubsection{Case 2: $v\perp B$}
The HQ velocity could also be in a plane perpendicular to $\bm{B}$ (\textit{i.e.} the $x$-$y$ plane). In such a case, one generally defines three momentum diffusion coefficients based on the direction of momentum transfer, as
\begin{align}
\kappa_{1}&=\int d^3 q \frac{d \Gamma(E,v)}{d^3 q} q_{x}^2.\label{k1}\\[0.2em]
\kappa_{2}&=\int d^3 q \frac{d \Gamma(E,v)}{d^3 q} q_y^2.\label{k2}\\[0.2em]
\kappa_{3}&=\int d^3 q \frac{d \Gamma(E,v)}{d^3 q} q_z^2.\label{k3}
\end{align} 
As we shall see, the structure of integrations will be different for the two cases. 
	
	\section{Perturbative determination of scattering rate $\Gamma$}
	As mentioned earlier, we consider coulomb scattering of the propagating heavy quark with the thermal quarks and gluons. To leading order, these $2\rightarrow 2$ processes are represented by the following tree level Feynman diagrams.
	\begin{figure}[H]
		\centering
		\includegraphics[scale=0.3]{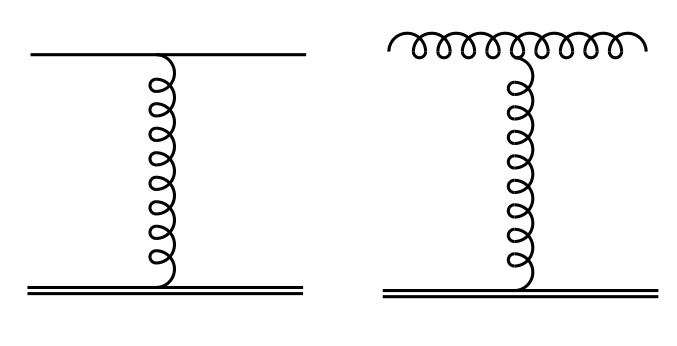}
		\caption{Feynman diagrams of processes contributing to heavy quark diffusion at leading order.}
		\label{coulomb}
	\end{figure} 
	The double line represents the heavy quark, whereas the thermal light quark is represented by the single line. $\Gamma$ calculated using the tree level diagrams in Fig.\eqref{coulomb} turns out to be quadratically infrared divergent which corroborates with the well known fact that the total rate of coulomb scattering in a plasma is quadratically infrared divergent\cite{Landau:1969}. Using a resummed gluon propagator in Fig.\eqref{coulomb} instead of a bare one softens the divergence to a logarithmic one\cite{Braaten:PRD44'1991}. This arises because the dynamical screening of the magnetic interaction provided by the transverse effective propagator is not sufficient to completely
	screen the divergence from the long-range static magnetic interaction. However, the two additional powers of $q$ in Eq.\eqref{kappa} render $\kappa$ infrared finite. The presence of the logarithm reflects that $\Gamma$ receives contribution from both the soft and hard momentum transfers. Soft processes involve $\bm{q}\sim gT$ and occur at a rate $\Gamma_{\text{soft}}\sim g^2T$, whereas the relatively scarce hard processes correspond to $\bm{q}\sim T$ and occur at a rate $\Gamma_{\text{hard}}\sim g^4T$. In this article, we shall be evaluating the soft contribution to $\Gamma$, and therefore, to the heavy quark diffusion coefficient, since it dominates over hard processes.
	
\begin{figure}[H]
		\centering
		\includegraphics[scale=0.40]{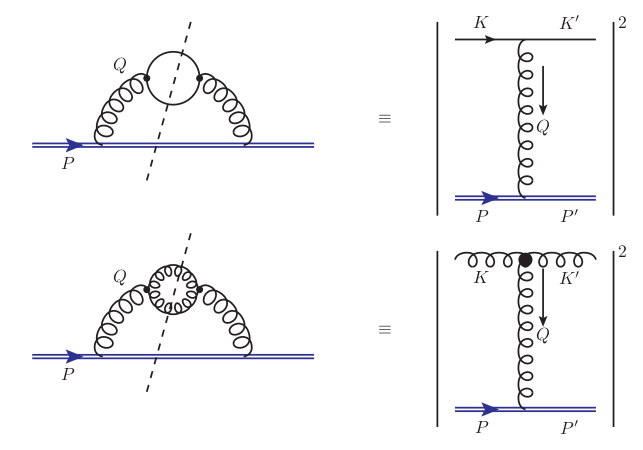}
		\caption{Cut (imaginary) part of heavy quark self energy diagrams yield the amplitude squared of $t$ channel scattering processes $qH\rightarrow qH$ and $gH\rightarrow gH$}
		\label{cutting}
	\end{figure}

\begin{figure}[H]
	\centering
	\includegraphics[scale=1.1]{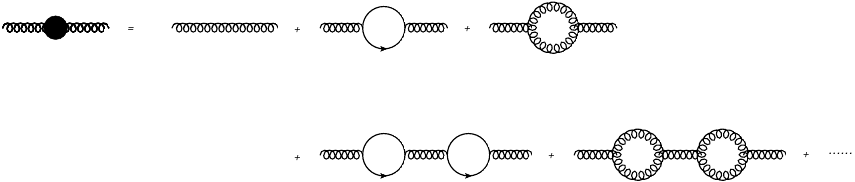}
	\caption{ Resummed gluon propagator. In addition to the leading order diagrams, resummation takes into account all higher order diagrams that contribute to leading order in $g$.}
	\label{resum}
\end{figure}

\begin{figure}[H]
	\centering
	\includegraphics[scale=0.3]{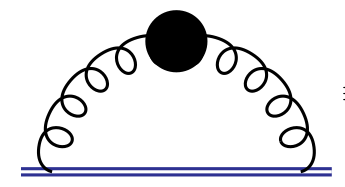}
	\caption{ HQ self-energy diagram with resummed gluon propagator.}
	\label{diagram}
\end{figure}

	An efficient method of calculating the scattering rate was put forward by Weldon\cite{Weldon:PRD28'2007} wherein, $\Gamma$ is evaluated from the imaginary part of the heavy quark self energy:
	\begin{equation}
		\Gamma(P \equiv E, \mathbf{v})=-\frac{1}{2 E} [1-n_F(E)] \operatorname{Tr}\left[(\not P+M_Q) \operatorname{Im} \Sigma\left(p_0+i \epsilon, \bm{p}\right)\right].\label{gamma}
	\end{equation}
	The imaginary part of the heavy quark self energy is related to the squared amplitude for coulomb scattering processes via the cutting rules, for the 2 loop self energy diagrams shown in Fig.\eqref{cutting}. This procedure automatically rules out using one-loop self energy diagrams, since the cut (imaginary) parts of those diagrams correspond to processes which do not conserve energy-momentum and thus are unphysical\cite{Thoma:arxiv}.
	The hard contribution to $\Gamma$ comes from the two loop self energy diagrams of Fig.\eqref{cutting}. However, when the gluon momentum is soft,  hard thermal loop corrections to the gluon propagator contribute at leading order in the strong coupling $g$ and therefore must be resummed. The resummed propagator is obtained by summing all possible self-energy corrections proportional to $g^2T^2$ to the bare propagator, as shown in Fig.\eqref{resum}. This is a geometric series summation, where the second and third diagrams are nominally of order $g^2T^2$. Similarly, the the fourth and fifth diagrams are $\mathcal{O}(g^4T^4)$, and so on\cite{Braaten:PRD44'1991}.
The self-energy diagram therefore to be evaluated  is the heavy quark self energy with resummed gluon propagator [Fig.\eqref{diagram}]. 
	We use the imaginary time formalism to compute the heavy quark self energy $\Sigma(P)$. Using Feynman rules,  $\Sigma(P)$ in a weak background magnetic field is given by:
	\begin{equation}
		\Sigma(P)=i g^2 \int \frac{d^4 Q}{(2 \pi)^4} \mathcal{D}^{\mu \nu}(Q) \gamma_\mu S(P-Q) \gamma_\nu.\label{self_energy}
	\end{equation}
	Since we work in the regime $\frac{qB}{M_Q}\ll 1$, we can ignore Landau quantization of the heavy quark energy levels, as has been done in\cite{Singh:JHEP68'2020}, and write the heavy quark propagator as
	\begin{equation}
		iS(P-Q\equiv K)=i\frac{\slashed{K}+M_Q}{K^2-M_Q^2}.
	\end{equation}
	The effective gluon propagator in the presence of a weak magnetic field is expressed as\cite{Karmakar:EPJC79'2019}
	\begin{equation}\label{dmunu}
		\begin{aligned}
			\mathcal{D}^{\mu \nu}(Q)= & \frac{\xi Q^\mu Q^\nu}{Q^4}+\frac{\left(Q^2-d\right) \Delta_1^{\mu \nu}}{\left(Q^2-b\right)\left(Q^2-d\right)-a^2}+\frac{\Delta_2^{\mu \nu}}{Q^2-c} \\
			& +\frac{\left(Q^2-b\right) \Delta_3^{\mu \nu}}{\left(Q^2-b\right)\left(Q^2-d\right)-a^2}+\frac{a \Delta_4^{\mu \nu}}{\left(Q^2-b\right)\left(Q^2-d\right)-a^2},
		\end{aligned}
	\end{equation}
	where,
	\begin{subequations}\label{forms}
		\begin{align}
			b(Q)&=\Delta_1^{\mu\nu}\Pi_{\mu\nu}\\
			c(Q)&=\Delta_2^{\mu\nu}\Pi_{\mu\nu}\\
			d(Q)&=\Delta_3^{\mu\nu}\Pi_{\mu\nu}\\
			a(Q)&=\frac{1}{2}\Delta_4^{\mu\nu}\Pi_{\mu\nu}
		\end{align}
	\end{subequations}
	$\Pi_{\mu\nu}(Q)$ is the gluon self energy computed within the HTL approximation. $\Delta_i^{\mu\nu}$ are the projection tensors along which the gluon self energy and the effective gluon propagator are expressed in the presence of a finite magnetic field, details of which can be found in Appendix \ref{AppendixA}. They are expressed as
	\begin{subequations}\label{pro}
		\begin{align}
			\Delta_1^{\mu \nu} & =\frac{1}{\bar{u}^2} \bar{u}^\mu \bar{u}^\nu, \\
			\Delta_2^{\mu \nu} & =g_{\perp}^{\mu \nu}-\frac{Q_{\perp}^\mu Q_{\perp}^\nu}{Q_{\perp}^2}, \\
			\Delta_3^{\mu \nu} & =\frac{\bar{n}^\mu \bar{n}^\nu}{\bar{n}^2}, \\
			\Delta_4^{\mu \nu} & =\frac{\bar{u}^\mu \bar{n}^\nu+\bar{u}^\nu \bar{n}^\mu}{\sqrt{\bar{u}^2} \sqrt{\bar{n}^2}}.
		\end{align}
	\end{subequations}
	$u^{\mu}$ is the velocity of the heat bath and $n^{\mu}$ can be considered to define the direction of the background magnetic field. Evaluating the form factors $b(Q)$, $c(Q)$, $d(Q)$, $a(Q)$ of Eqs.\eqref{forms} is akin to evaluating the effective gluon propagator. The calculation of these form factors under HTL approximation, along with the other tensors in Eq.\eqref{pro} is detailed in Appendix \ref{AppendixB}. 
	
	Following\cite{Bandyopadhyay:PRD105'2022}, we next evaluate the trace in Eq.\eqref{gamma}
	\begin{equation}\label{trace}
		\begin{aligned}
			\operatorname{Tr}[(\slashed{P}+M_Q) \Sigma(P)]= & ig^2 \int \frac{d^4 Q}{(2 \pi)^4} \frac{1}{K^2-M_Q^2} \\
			& \times \sum_{i=1}^4 \chi_i \operatorname{Tr}\left[(\slashed{P}+M_Q) \Delta_i^{\mu \nu} \gamma_\mu\left(\slashed{K}+M_Q\right) \gamma_\nu\right]
		\end{aligned}
	\end{equation}
	Taking the gauge parameter $\xi$ in Eq.\eqref{dmunu} to be 0, the coefficients $\chi_i$'s are given by:
	\begin{subequations}
		\begin{align}
			\chi_1 & =\frac{\left(Q^2-d\right)}{\left(Q^2-b\right)\left(Q^2-d\right)-a^2}, \\
			\chi_2 & =\frac{1}{\left(Q^2-c\right)}, \\
			\chi_3 & =\frac{\left(Q^2-b\right)}{\left(Q^2-b\right)\left(Q^2-d\right)-a^2}, \\
			\chi_4 & =\frac{a}{\left(Q^2-b\right)\left(Q^2-d\right)-a^2} .
		\end{align}
	\end{subequations}
	 It should be noted that the gauge can be fixed only when the quantity being calculated is known to be gauge invariant. In our case, the scattering rate $\Gamma$ is known to be gauge invariant in QED as well as in QCD, as described in \cite{Braaten:PRD44'1991}. In fact the soft and hard contributions to $\Gamma$ are separately gauge invariant and can be calculated in any suitable gauge. In our case, we set $\xi=0$, which corresponds to the Landau gauge. 
	
	We evaluate the individual traces in Eq.\eqref{trace}.
	\begin{subequations}
		\begin{align}
			\operatorname{Tr}\left[(\slashed{P}+M_Q) \Delta_1^{\mu \nu} \gamma_\mu\left(\slashed{K}+M_Q\right) \gamma_\nu\right]&=\frac{4}{\bar{u}^2}\left[2(P.\bar{u})(K.\bar{u})+\bar{u}^2(M^2-P.K)\right]\\
			&=A_1+B_1\nonumber\\[0.4em]
			\operatorname{Tr}\left[(\slashed{P}+M_Q) \Delta_2^{\mu \nu} \gamma_\mu\left(\slashed{K}+M_Q\right) \gamma_\nu\right]\!&=\!4\!\left[2(P.K)_{\perp}\!-\!\frac{2(P.Q)\!_{\perp}\!(K.Q)_{\perp}}{Q_{\perp}^2}\!+\!(M^2\!-\!P.K)\right]\\
			&=A_2+B_2\nonumber\\[0.4em]
			\operatorname{Tr}\left[(\slashed{P}+M_Q) \Delta_3^{\mu \nu} \gamma_\mu\left(\slashed{K}+M_Q\right) \gamma_\nu\right]&=\frac{4}{\bar{n}^2}\left[2(P.\bar{n})(K.\bar{n})+\bar{n}^2(M^2-P.K)\right]\\&=A_3+B_3\nonumber\\[0.4em]
			\operatorname{Tr}\left[(\slashed{P}+M_Q) \Delta_4^{\mu \nu} \gamma_\mu\left(\slashed{K}+M_Q\right) \gamma_\nu\right]&=\frac{8}{\sqrt{\bar{n}^2}\sqrt{\bar{u}^2}}\left[(P.\bar{u})(K.\bar{n})+(P.\bar{n})(K.\bar{u})+(\bar{u}\cdot\bar{n})(M^2-P.K)\right]\nonumber\\
			&=A_4+B_4
		\end{align}
	\end{subequations}
	The traces have been separated into $q_0$ independent and $q_0$ dependent terms denoted by $A_i$ and $B_i$ respectively; $i=1,2,3,4$. This is done to facilitate the frequency sum over $q_0$, as will be seen later. The condition required for such a separation to be executed is that the transfer momentum four-vector $Q^{\mu}$ be spacelike, which is indeed the case for $t$ channel scattering processes. The $A_i'$s and $B_i'$s come out to be
	\begin{equation}\label{A1}
		A_1=4(2p_0^2-\bm{p}\cdot\bm{q}),\quad B_1=-8p_0^2\frac{q_0^2}{q^2}-8\frac{q_0^2(P.Q)^2}{Q^2q^2}+16p_0q_0\frac{P\cdot Q}{q^2}
	\end{equation}
	
	\vspace{-6mm}
	\begin{equation}\label{A2}
		A_2=4\left[2\{P_{\perp}^2-(P\cdot Q)_{\perp}\}-\bm{p}\cdot \bm{q} +\frac{2(P.Q)_{\perp}\{Q_{\perp}^2-(P.Q)_{\perp}\}}{Q_{\perp}^2}\right],\quad B_2=4p_0q_0
	\end{equation}
	
	\vspace{-2mm}
	\begin{equation}\label{A3}
		A_3 =\frac{8}{\bar{n}^2}\left[p_3^2-\frac{2 p_3 q_3}{q^2}(\bm{p} \cdot \bm{q}) +\frac{q_3}{q^2}(\bm{p} \cdot \bm{q})^2-\frac{\bar{n}^2}{2}(\bm{p} \cdot \bm{q})\right],\quad B_3 =4 p_0 q_0
	\end{equation}
	
	\vspace{-2mm}
	\begin{equation}\label{A4}
		A_4 =\frac{16}{\sqrt{\bar{n}^2}}\left[-p_0 p_3+\frac{p_0 p_3}{q^2}(\bm{p} \cdot \bm{q})\right]=\frac{16 p_0 p_3}{\sqrt{\bar{n}^2}}\left[\frac{\bm{p} \cdot \bm{q}}{q^2}-1\right]
	\end{equation}
	\begin{align}
		B_4 &=\frac{16}{\sqrt{\bar{n}^2}}\left[\frac{p_3 q_0^2 p_0}{q^2}-\frac{p_3 q_0}{q^2}(\bm{p} \cdot \bm{q})-\frac{q_0^2 q_3 p_0}{Q^2 q^2}(\bm{p} \cdot \bm{q})+\frac{q_0 q_3}{Q^2 q^2}(p \cdot \bm{q})^2\right] \nonumber\\
		& \times\left\{\left(-\frac{q_0^2}{q^2}\right)+\text { higher powers of } \frac{q_0^2}{q^2}\right\}
	\end{align}
	Next, we perform the frequency sum over $q_0$. To that end, a convenient method is to introduce spectral representations for the propagators\cite{Pisarski:NPB309'1988}. The fermion propagator is spectrally represented as
	\begin{equation}
		\frac{1}{K^2-M_Q^2}=-\frac{1}{2E'}\int_{0}^{\beta}d\tau'e^{k_0\tau'}\left[(1-n_F(E'))e^{-E'\tau'}-n_F(E')e^{E'\tau'}\right],
	\end{equation}
	where, $E'=\sqrt{k^2+M_Q^2}$, $\beta=1/T$ Similarly, pieces of the effective gluon propagator $\chi_i$ can be expressed as
	\begin{equation}
		\chi_i=-\int_{0}^{\beta}d\tau\,e^{q_0\tau}\int_{-\infty}^{\infty}d\omega\,\rho_i(\omega,q)\left[1+n_B(\omega)\right]e^{-\omega\tau}.
	\end{equation}
	$\rho_i$ are the spectral functions associated with $\chi_i$, and are odd functions of $\omega$. Each spectral function contains contributions from both spacelike and timelike frequencies, and is expressed as 
	\begin{equation}\label{cutpole}
		\rho_i(\omega,q)=\rho_i^{\text{pole}}(\omega,q)+\rho_i^{\text{cut}}(\omega,q),
	\end{equation}
	with
	\begin{align}
		\rho_i^{\text{pole}}(\omega,q)&=\rho_i^{\text{res}}\delta(\omega-\omega_i(q))\\
		\rho_i^{\text{cut}}(\omega,q)&=\rho_i^{\text{dis}}\theta(q^2-\omega^2)
	\end{align}
	Thus, the spectral functions have delta function contributions at the timelike points (poles) $\omega=\omega_i(q)$, where, $\omega_i(q)$ are the dispersion relations, and $\rho_i^{\text{res}}$ are the residues at those points. For spacelike frequencies $|\omega|<q$, $\rho_i'$s receive a discontinuous contribution from the imaginary part of the resummed propagator (Landau damping) 
	\begin{equation}
		\rho_i^{\text{dis}}(\omega,q)=-\frac{1}{\pi}\text{Im}\left(\chi_i\Big|_{q_0=\omega+i\epsilon}\right).
	\end{equation}
	Since we work in the regime $|\omega|<q$, only the cut part in Eq.\eqref{cutpole} contributes and is denoted simply as $\rho$ hereafter.  
	The calculation and final expressions of the $\rho_i$'s are given in Appendix \ref{AppendixC}.
	The advantage of the spectral function representation is that it simplifies the evaluation of the frequency sums due to the appearance of delta functions in the integral, coming from
	\begin{subequations}
		\begin{align}
			T\sum_{q_0}e^{q_0(\tau-\tau')}=\delta(\tau-\tau')\label{delta1}\\
			T\sum_{q_0}q_0\,e^{q_0(\tau-\tau')}=\delta'(\tau-\tau')\label{delta2}
		\end{align}
	\end{subequations}
	Using this, Eq.\eqref{trace} becomes
	\begin{align}
		&\operatorname{Tr}[(\not P+M) \Sigma(P)]\nonumber \\
		&=-ig^2 \int \frac{d^4 Q}{(2 \pi)^4} \frac{1}{K^2-M_Q^2}\,\sum_{i=1}^4\chi_i[A_i+B_i]\nonumber \\[0.2em]
		&=-g^2 T \sum_{i=1}^4 \int \frac{d^3 q}{(2 \pi)^3}\int_{-\infty}^{+\infty} d \omega\left[1+n_B(\omega)\right] \int_0^{\beta} d \tau^{\prime} \int_0^{\beta} d \tau e^{p_0 \tau^{\prime}}e^{-\omega\tau} \nonumber\\
		&\times \sum_{q_0} e^{q_0\left(\tau-\tau^{\prime}\right)}\left[A_i+B_i\right] \frac{\rho_i(\omega, q)}{2 E^{\prime}}\left[\left\{1-n_F(E')\right\} e^{-E^{\prime} \tau^{\prime}}-n_F\left(E^{\prime}\right) e^{E'}\tau^{\prime}\right]\nonumber\\
		&=-g^2 T \sum_{i=1}^4 \int \frac{d^3 q}{(2 \pi)^3}\int_{-\infty}^{+\infty} d \omega\left[1+n_B(\omega)\right]\left(I_1+I_2\right),\label{trace2}
	\end{align}
	where, using Eq.\eqref{delta1}, 
	\begin{align}
		I_1&= \int_0^{\beta} d \tau^{\prime} \int_0^{\beta} d \tau\, e^{p_0 \tau^{\prime}}e^{-\omega\tau} 
		A_i\,\delta(\tau-\tau') \frac{\rho_i(\omega, q)}{2 E^{\prime}}\nonumber\\&\times\left[\left\{1-n_F(E')\right\} e^{-E^{\prime} \tau^{\prime}}-n_F\left(E^{\prime}\right) e^{E'}\tau^{\prime}\right].
	\end{align}
	We use the $\delta$ function to integrate over $\tau'$ to obtain
	\begin{align}
		I_1=\int_{0}^{\beta}d\tau e^{(p_0-\omega)\tau}
		\frac{\rho_i(\omega, q)}{2 E^{\prime}}\left[\left\{1-n_F(E')\right\} e^{-E^{\prime} \tau}-n_F\left(E^{\prime}\right) e^{E'\tau}\right]A_i.
	\end{align}
	The $\tau$ integration 
	ultimately yields
	\begin{equation}
		I_1=-\sum_{j=\pm 1}\frac{j\,n_F(jE')}{p_0-\omega+jE'}\left[e^{(p_0-\omega+jE')\beta}-1\right]A_i.\label{i1}
	\end{equation}
	Since, $B_i$ is $q_0$ dependent, a sample term can be written as $B_i=q_0C_i$. Then, using Eq.\eqref{delta2} yields 
	\begin{align}
		I_2&= \int_0^{\beta} d \tau^{\prime} \int_0^{\beta} d \tau\, e^{p_0 \tau^{\prime}}e^{-\omega\tau} 
		\sum_{i}C_i\,\delta'(\tau-\tau') \frac{\rho_i(\omega, q)}{2 E^{\prime}}\left[\left\{1-n_F(E')\right\} e^{-E^{\prime} \tau^{\prime}}-n_F\left(E^{\prime}\right) e^{E'\tau'}\right]\nonumber\\[0.2em]
		&=-\int_{0}^{\beta}d\tau \frac{d}{d\tau}e^{(p_0-\omega)\tau}\left[\left\{1-n_F(E')\right\} e^{-E^{\prime} \tau}-n_F\left(E^{\prime}\right) e^{E'\tau}\right]C_i\nonumber\\[0.3em]
		&=\sum_{j=\pm 1}j\,n_F(jE')\left[e^{(p_o-\omega+jE')\beta}-1\right]C_i.\label{i2}
	\end{align}
	$p_0$ is discrete since we are working in the imaginary time formalism. Specifically, $p_0=i(2n+1)\pi/\beta$. At these discrete energies, $e^{p_0\beta}=-1$ and $p_0$ thus gets eliminated from the exponent in Eqs.\eqref{i1} and \eqref{i2}. Thereafter, we analytically continue $p_0$ to real values via $p_0\rightarrow E+i\omega$. The imaginary part is then extracted, which comes from energy denominator terms of the form
	\begin{equation}
		\text{Im}\left(\frac{1}{p_0+E'-\omega}\right)\Big|_{p_0\rightarrow E+i\omega}=-i\pi\delta(p_0+E'-\omega).\label{im}
	\end{equation}
	Since there is no energy denominator in Eq.\eqref{i2}, $I_2$ does not have any imaginary part, and the contribution to the imaginary part of the self energy thus comes solely from $I_1$. Using Eq.\eqref{trace2}, \eqref{i1} and \eqref{im}, we can write
	\begin{align}
		&\operatorname{Tr}\left[(\not P+M_Q) \operatorname{Im} \Sigma\left(p_0+i \epsilon, \bm{p}\right)\right]\nonumber\\[0.3em]
		= & \pi g^2  \sum_{i=1}^4 \int \frac{d^3 q}{(2 \pi)^3}  \int_{-\infty}^{\infty} d \omega\left[1+n_B(\omega)\right] \frac{\rho_i(\omega, q) A_i}{2 E'}\nonumber \\
		& \times \sum_{j= \pm 1} j n_F\left(\sigma E'\right)\left(e^{\left(\sigma E'-\omega\right)\beta}+1\right) \delta\left(E+j E'-\omega\right)\nonumber \\
		= & \pi g^2 \left(e^{-E\beta}+1\right) \sum_{i=1}^4 \int \frac{d^3 q}{(2 \pi)^3}  \int_{-\infty}^{+\infty} d \omega\left[1+n_B(\omega)\right] \frac{\rho_i(\omega, q) A_i}{2 E'} \nonumber\\
		& \times \sum_{j=\pm1} j n_F\left(j E'\right) \delta\left(E+j E'-\omega\right) .
	\end{align}
	Thus, using all the results, $\Gamma$ in Eq.\eqref{gamma} is given by
	\begin{align}
		\Gamma(E,\bm{v})&=-\frac{\pi g^2}{2E}\sum_{i=1}^4\int\frac{d^3 q}{(2 \pi)^3}  \int_{-\infty}^{+\infty} d \omega\left[1+n_B(\omega)\right] \frac{\rho_i(\omega, q) A_i}{2 E'}\nonumber \\
		& \times \sum_{j=\pm1} j n_F\left(j E'\right) \delta\left(E+j E'-\omega\right).\label{gamma1}
	\end{align}
	We now simplify the above expression further by recalling that $M_Q$, $p$ $\gg T.$  The delta function corresponding to $j=1$ does not contribute for $\omega \leq T$, and so can be dropped. For $E'\gg T$, the fermi distribution function is exponentially suppressed, so that $n_F(E')\approx 0$. Employing  these approximations, we have
	\begin{equation}
		\Gamma(E,\bm{v})=\frac{\pi g^2}{2E}\sum_{i=1}^4\int\frac{d^3 q}{(2 \pi)^3}  \int_{-\infty}^{+\infty} d \omega\left[1+n_B(\omega)\right] \frac{\rho_i(\omega, q) A_i}{2 E'}\delta(E-E'-\omega).\label{gamma2}
	\end{equation}
Further, we have
	\begin{align}
	E'&=\sqrt{(\bm{p}-\bm{q})^2+M_Q^2}\nonumber\\
	&\simeq E\left(1-\frac{2\bm{p}\cdot \bm{q}}{E^2}\right)^{1/2}\nonumber\\
	&\simeq E-\bm{v}\cdot \bm{q}.\label{approx}
\end{align}
Although $E-E'\sim \mathcal{O}(v)$, $\frac{1}{E}-\frac{1}{E'}\sim \mathcal{O}(v^2)$, which we neglect. So, $\frac{1}{E}\approx\frac{1}{E'}$. Thus, we obtain
\begin{equation}
	\Gamma(E,\bm{v})=\frac{\pi g^2}{2E}\sum_{i=1}^4\int\frac{d^3 q}{(2 \pi)^3}  \int_{-\infty}^{+\infty} d \omega\left[1+n_B(\omega)\right]\frac{\rho_i(\omega, q) A_i}{2 E}\delta(\omega-\bm{v}\cdot\bm{q}).\label{gamma_final}
\end{equation}
	
	\section{Energy loss and momentum diffusion coefficients}
	\subsection{Case 1: $v\parallel B$}
	After having computed $\Gamma$, we use it to evaluate dynamic quantities such as the heavy quark energy loss and the momentum diffusion coefficient. The HQ velocity points along $\bm{B}$, which, in turn is taken to be of constant
	magnitude, and pointing along $\hat{z}$. 	The energy loss of the heavy quark propagating through the high temperature QCD plasma is given by Eq.\eqref{dedx}. Using Eq.\eqref{gamma2} and the approximations mentioned above, we get, for the energy loss:
	\begin{equation}
		\frac{dE}{dx}=\frac{\pi g^2}{2Ev}\sum_{i=1}^4\int\frac{d^3 q}{(2 \pi)^3}  \int_{-\infty}^{+\infty} d \omega\left[1+n_B(\omega)\right]\omega\, \frac{\rho_i(\omega, q) A_i^{\parallel}}{2 E}\delta(\omega-\bm{v}\cdot\bm{q}),\label{dedx2}
	\end{equation}
where, $A_i^{\parallel}$ are the $A_i$ from Eqs.(\ref{A1}-\ref{A4}) evaluated with $v_{\perp}=0$.	For $\omega\ll T$, the  Bose distribution function can be written as an expansion in $\omega/T$ so that
	\begin{equation}
		1+n_B(\omega)\simeq \frac{T}{\omega}+\frac{1}{2}-\mathcal{O}\left(\frac{\omega}{T}\right)+\mathcal{O}\left(\frac{\omega}{T}\right)^2-\cdots\label{be}
	\end{equation}
	Now, the $\rho_i'$s in Eq.\eqref{dedx2} are odd functions of $\omega$\cite{Pisarski:NPA498'1989}. Hence, only the even part of $1+n_B(\omega)$ will contribute to the integral, since the integration over $\omega$ is symmetric. Thus, we have
	\begin{equation}
		\frac{dE}{dx}=\frac{\pi g^2}{8E^2v}\sum_{i=1}^4\int\frac{d^3 q}{(2 \pi)^3}  \int_{-\infty}^{+\infty} d \omega\,\omega\, \rho_i(\omega, q) A_i^{\parallel}\,\delta(\omega-\bm{v}\cdot\bm{q}).\label{dedx3}
	\end{equation} 
	The momentum diffusion coefficients are given by
	\begin{align}
		\kappa_L&=\frac{\pi g^2}{2E}\sum_{i=1}^4\int\frac{d^3 q}{(2 \pi)^3} q_L^2 \int_{-\infty}^{+\infty} d \omega\left[1+n_B(\omega)\right] \frac{\rho_i(\omega, q) A_i^{\parallel}}{2 E}\delta(\omega-\bm{v}\cdot\bm{q}).\label{kappa2L}\\
		\kappa_T&=\frac{\pi g^2}{2E}\sum_{i=1}^4\int\frac{d^3 q}{(2 \pi)^3} q_T^2 \int_{-\infty}^{+\infty} d \omega\left[1+n_B(\omega)\right] \frac{\rho_i(\omega, q) A_i^{\parallel}}{2 E}\delta(\omega-\bm{v}\cdot\bm{q}).\label{kappa2T}
	\end{align}
	This time, only the odd part of $1+n_B(\omega)$ will contribute to the integral. Thus, we have
	\begin{align}
		\kappa_L&=\frac{\pi g^2T}{4E^2}\sum_{i=1}^4\int\frac{d^3 q}{(2 \pi)^3} q_L^2 \int_{-\infty}^{+\infty} d \omega\frac{\rho_i(\omega, q) A_i^{\parallel}}{\omega}\delta(\omega-\bm{v}\cdot\bm{q}).\label{kappa3L}\\
		\kappa_T&=\frac{\pi g^2T}{4E^2}\sum_{i=1}^4\int\frac{d^3 q}{(2 \pi)^3} q_T^2 \int_{-\infty}^{+\infty} d \omega\frac{\rho_i(\omega, q) A_i^{\parallel}}{\omega}\delta(\omega-\bm{v}\cdot\bm{q}).\label{kappa3T}
	\end{align}
	Because the HQ velocity points along $\bm{B}$ (hence, along $\hat{z}$), we have $\bm{v}\cdot\bm{q}=vq \cos \theta\equiv vq\eta$, where, $\theta$ is both the angle between $\bm{q}$ and the $\bm{v}$, as well as the polar angle of integration. The delta function is then used to integrate over $\eta$ with $d^3q=2\pi q^2dqd\eta$, which sets $\omega=vq\eta$. Since $-1\leq \eta\leq 1$, $-vq\leq \omega\leq vq$. This finally leads to
	\begin{align}
		\frac{dE}{dx}&=\frac{\pi g^2}{8E^2\,v^2(2\pi)^2}\int dq\, q\int_{-vq}^{vq} d \omega\,\omega\, \sum_{i=1}^4\rho_i(\omega, q,\frac{\omega}{vq}) A_i^{\parallel}.\label{dedx4}\\
		\kappa_L&=\frac{\pi g^2T}{4E^2\,v}\int dq\, q^3\int_{-vq}^{vq} d \omega\sum_{i=1}^4\frac{\rho_i(\omega, q,\frac{\omega}{vq}) A_i^{\parallel}}{\omega}\left(\frac{\omega^2}{v^2q^2}\right).\label{kappa4L}\\
		\kappa_T&=\frac{\pi g^2T}{4E^2\,v}\int dq\, q^3\int_{-vq}^{vq} d \omega\sum_{i=1}^4\frac{\rho_i(\omega, q,\frac{\omega}{vq}) A_i^{\parallel}}{\omega}\left(1-\frac{\omega^2}{v^2q^2}\right).\label{kappa4T}
	\end{align}

\subsection{Case 2: $v\perp B$}
 The HQ velocity now lies in the $x$-$y$ plane, with the magnetic field pointing in the $\hat{z}$ direction as earlier. Since $\bm{v}$ is no longer oriented along the $z$ axis, $\bm{v}\cdot\bm{q}$ is not trivial. The direction of HQ velocity in the $x$-$y$ plane can be specified by the azimuthal angle $\phi'$ (The polar angle $\theta'=0$, as it is in the $x$-$y$ plane). The vector $\bm{q}$ is specified by $q$, $\theta$ and $\phi$ (our integration variables), where, $\theta$ is the polar angle, and $\phi$ is the azimuthal angle. Then,
\begin{equation}
	\bm{v}\cdot\bm{q}=vq\sin\theta\cos(\phi-\phi'),
\end{equation}
where, $v=|\bm{v}|$, $q=|\bm{q}|$. The interaction rate becomes\footnote{The limits of the $q$ integration is discussed in the next section.}
	\begin{align}
			\Gamma(E,\bm{v})=\frac{\pi g^2}{4E^2(2\pi)3}\sum_{i=1}^4&\int dq\int_0^\pi d\theta\sin\theta\int_0^{2\pi}d\phi  \int_{-\infty}^{+\infty} d \omega\left[1+n_B(\omega)\right]\\\nonumber
			&\rho_i(\omega, q) A_i^{\perp}\delta[\omega-vq\sin\theta\cos(\phi-\phi')],\label{gamma_perp}
	\end{align}	
where, $A_i^{\perp}$ are the $A_i$ from Eqs.(\ref{A1}-\ref{A4}) evaluated with $v_z=0$. We introduce a variable $y=\phi-\phi'$. $\Gamma$ then simplifies to
\begin{multline}
	\Gamma(E,\bm{v})=\frac{g^2}{32E^2v\pi^2}\int dq\,q\int_0^\pi d\theta \sin\theta\,\delta\left(\sin\theta-\frac{\omega}{vq\cos y}\right)\times\\\int_{-\phi'}^{2\pi-\phi'}\frac{dy}{\cos y}\int_{-\infty}^{+\infty} d \omega\left[1+n_B(\omega)\right]\sum_{i=1}^4A_i^{\perp}\rho_i(\omega, q).
\end{multline} 
We use the result
\begin{equation}
\int_0^\pi d\theta \sin\theta\,\delta\left(\sin\theta-c\right)=\frac{2c}{\sqrt{1-c^2}}\Theta(c)\Theta(1-c)\,;\,c\in \mathbb{R},
\end{equation}
to integrate over $\theta$. The $\Theta$ function sets 
\begin{align}
	&0\leq \frac{\omega}{v q\cos y}\leq 1 \nonumber\\ 
	&0\leq \omega\leq vq\cos y
\end{align} $\Gamma$ then finally becomes
\begin{multline}
	\Gamma(E,\bm{v})= \frac{g^2}{16E^2v\pi^2}\int dq\int_{-\phi'}^{2\pi-\phi'}dy\int_0^{v q\cos y} d \omega\\ \frac{q}{\sqrt{v^2q^2\cos^2y/\omega^2-1}} \frac{\left[1+n_B(\omega)\right]}{\cos y}\sum_{i=1}^4A_i^{\perp}\rho_i(\omega, q).
\end{multline}
From Eqs. (\ref{k1}-\ref{k3}), the momentum diffusion coefficients are  obtained as: 
\begin{align}
	\kappa_{1}&=\frac{g^2}{16E^2v^3\pi^2}\int dq\int_{-\phi'}^{2\pi-\phi'}dy\int_0^{v q\cos y} d \omega\,
	\frac{\omega^3\cos^2(y+\phi')[1+n_B(\omega)]}{\cos^3y\sqrt{v^2q^2\cos^2y-\omega^2}}\sum_{i=1}^4A_i^{\perp}\rho_i(\omega, q).\label{k1_perp}\\[0.3em]
\kappa_{2}&=\frac{g^2}{16E^2v^3\pi^2}\int dq\int_{-\phi'}^{2\pi-\phi'}dy\int_0^{v q\cos y} d \omega\,
\frac{\omega^3\sin^2(y+\phi')[1+n_B(\omega)]}{\cos^3y\sqrt{v^2q^2\cos^2y-\omega^2}}\sum_{i=1}^4A_i^{\perp}\rho_i(\omega, q).\label{k2_perp}\\[0.3em]
\kappa_{3}&=\frac{g^2}{16E^2v^3\pi^2}\int dq\int_{-\phi'}^{2\pi-\phi'}dy\int_0^{v q\cos y} d \omega\,
\frac{q\omega\sqrt{v^2q^2\cos^2y-\omega^2}[1+n_B(\omega)]}{\cos^3y}\sum_{i=1}^4A_i^{\perp}\rho_i(\omega, q).
\end{align}
Also, the HQ energy loss [from Eq.\eqref{dedx}] comes out to be
\begin{align}
	dE/dx=\frac{g^2}{16E^2v^2\pi^2}\int dq\int_{-\phi'}^{2\pi-\phi'}dy\int_0^{v q\cos y} d \omega\,\frac{\omega^2q[1+n_B(\omega)]}{\cos y\sqrt{v^2q^2\cos^2y-\omega^2}}\sum_{i=1}^4A_i^{\perp}\rho_i(\omega, q).
\end{align}

	\section{Results and Discussions}
	In this section, we present the results of the heavy quark (Charm and Bottom) momentum diffusion coefficients and the heavy quark energy loss. The running coupling constant is taken up to one-loop:
	\begin{equation}
		g(\Lambda)=\left[\frac{48 \pi^2}{(33-2N_f)\text{ln}\left(\frac{\Lambda^2}{\Lambda^2_{\overline{MS}}}\right)}\right]^{1/2}
	\end{equation}
	The renormalisation scale $\Lambda$ can be taken to be $2\pi T$ to introduce temperature dependence in the coupling. The $\overline{\text{MS}}$ scale is taken to be 176 MeV\cite{Bazavov:PRD86'2012}. The use of this form of the coupling is justified since $q_fB\ll T^2$. In the strong field limit, use of momentum-dependent couplings might be more appropriate\cite{Hattori:PRD95'2017}. The bottom and charm quark masses are taken to be 4.18 GeV and 1.28 GeV, respectively. The HQ momentum is taken to be $p=0.3$ GeV. Heavy quark dynamics with temperature dependent couplings was studied first in\cite{Das:PLB747'2015}.
	
	An important point to note is that the integrals in Eqs.\eqref{dedx4}, \eqref{kappa4L}, \eqref{kappa4T} are logarithmically U-V divergent and hence, require a U-V cut-off. Following the prescription of \cite{Beraudo:NPA831'2009}, we take the U-V cut-off to be $3.1Tg^{1/3}$. The reason for this divergence is that our calculations are confined to the region of soft gauge boson momentum transfer. In the $B=0$ case, it is shown explicitly that the dependence on this cut-off vanishes once the full range of momentum transfers is taken into account. We expect the same to be true in the case of weak magnetic fields too. However, the full calculation, including hard scatterings, is left for a future work. As mentioned earlier, the soft scatterings contribute to $\mathcal{O}(g(T)^2)$ in $\Gamma$ whereas the hard contribution to $\Gamma$ will be of $\mathcal{O}(g(T)^4)$. As such, it can be inferred that the major contribution to the momentum diffusion of the HQ via elastic scatterings comes from soft gluon exchange with the thermal quarks and gluons of the heat bath. It is also worth mentioning that this U-V cut-off is not necessary if one uses the Lowest Landau Level (LLL) approximation for the HQ propagator in the presence of a strong ($q_fB\gg T^2$) magnetic field, because of the presence of the exponential factor $e^{-k_{\perp}^2\big/|q_fB|}$ in the HQ propagator.
	
	\subsection{Case 1: $v\parallel B$}
	
	\begin{figure}[H]
		\begin{subfigure}{0.48\textwidth}
			\includegraphics[width=0.97\textwidth]{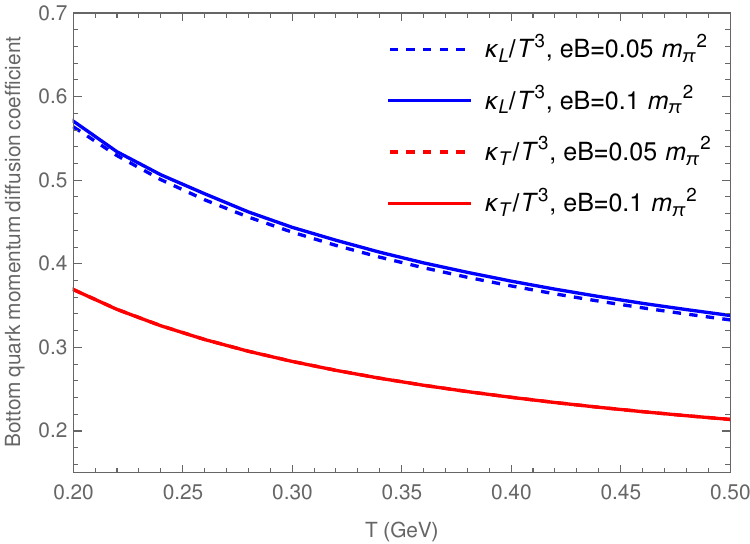}
			\caption{}\label{kTB}
		\end{subfigure}
		\hspace*{\fill}
		\begin{subfigure}{0.48\textwidth}
			\includegraphics[width=0.97\textwidth]{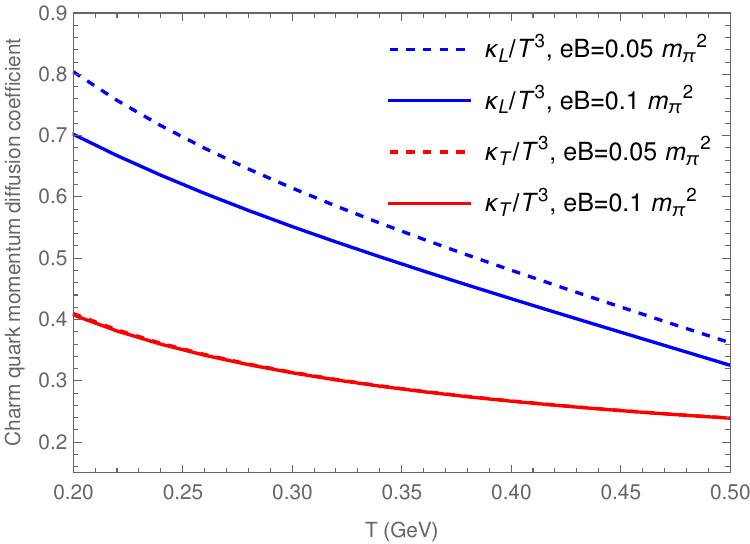}
			\caption{}\label{kLB}
		\end{subfigure}
		\caption{Normalised  momentum diffusion coefficients for Bottom (a) and Charm (b) quarks as a function of temperature at different fixed values of background magnetic field.}
		\label{kB}
	\end{figure}
	\begin{figure}[H]
		\centering
		\includegraphics[scale=0.65]{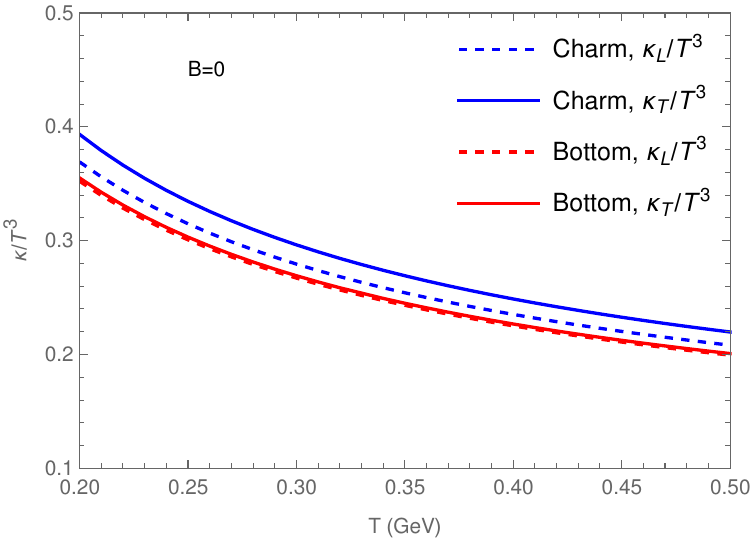}
		\caption{Normalised transverse and longitudinal momentum diffusion coefficients for both Charm and Bottom quarks as a function of temperature in the absence of a background magnetic field.}
		\label{kB0}
	\end{figure}
	
	The figures show the temperature variation of the momentum diffusion coefficients ($\kappa_{L/T}$) for heavy quarks at a fixed momentum [$p=0.3$ GeV]. For comparison, the $B=0$ results are also shown in Fig.\eqref{kB0}.  As can be seen from Figures \eqref{kTB} and \eqref{kLB}, both the longitudinal and transverse momentum diffusion coefficients show a monotonous decrease with temperature. For bottom quark, the magnitude of the longitudinal component is $\sim 1.5$ times larger than its transverse counterpart in the entire temperature range. For the charm quark, $\kappa_L/\kappa_T$ is much larger ($\sim 2$) at lower temperatures than at higher temperatures ($\sim 1.5$). The figures also show the behaviour of the coefficients with magnetic field. $\kappa_T$ (normalised) for both the heavy flavors show very little sensitivity to changes in magnetic field. In the case of $\kappa_L$, there is a strong dependence on $B$ strength for the Charm quark, which increases with decreasing field strength. For the bottom quark, however, both $\kappa_L$ and $\kappa_T$ decrease with decreasing $B$. Fig.\eqref{kB0} shows the $B=0$ result for the momentum diffusion coefficients. The momentum diffusion of Charm quarks is faster than that of the Bottom quark, owing to the smaller mass of the former. $\kappa_T$ is larger than $\kappa_L$ in the entire temperature range, for both the heavy flavours. Also, the degree of anisotropy is negligible in case of the bottom quark, whereas it is discernible in the case of charm quark. A large value of the momentum diffusion coefficient would work towards decreasing the yield of bound states such as the $J/\psi$ (charmonium), bottomonium, etc.; the Brownian motion of the heavy-quarks overwhelming the screened potential holding the $q$-$\bar{q}$ pair together. On the other hand, a stronger energy loss, $dE/dx$, of the propagating HQ would result in the stopping of a $q$-$\bar{q}$ pair (not so much for $b$-$\bar{b}$), leading to the increase in yield of mesonic bound states involving heavy quarks. The fate of a $q$-$\bar{q}$ pair produced in the initial stages of a heavy-ion collision thus depends on these competing factors. This phenomenon has been elucidated in detail in\cite{Svetitsky:PRD37'1988}
	
	\begin{figure}[H]
		\begin{subfigure}{0.48\textwidth}
			\includegraphics[width=0.95\textwidth]{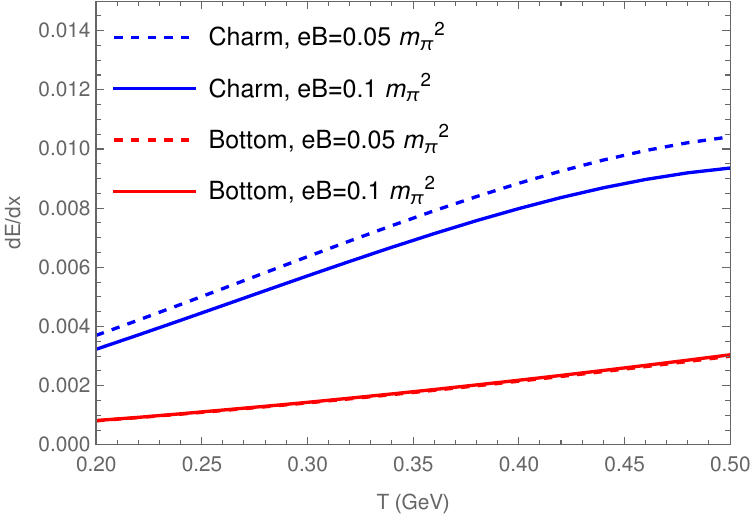}
			\caption{}\label{dedxB01}
		\end{subfigure}
		\hspace*{\fill}
		\begin{subfigure}{0.48\textwidth}
			\includegraphics[width=0.95\textwidth]{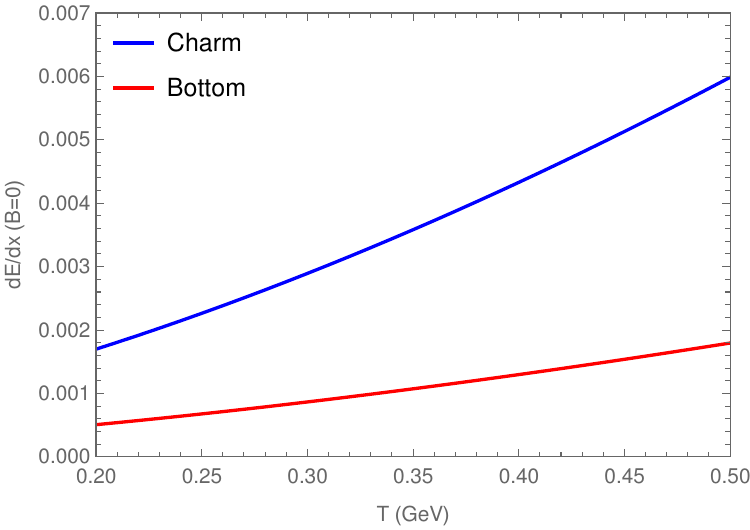}
			\caption{}\label{dedxB0}
		\end{subfigure}
		\caption{(a) Energy loss of heavy quark as a function of temperature in the presence of fixed values of background magnetic field of strengths . (b) Energy loss of heavy quarks in the absence of background magnetic field. }
		\label{figdedx}
	\end{figure}
	Fig.\eqref{dedxB01} shows the temperature variation of the HQ (Bottom and Charm) energy loss in the presence of a weak constant background field. In contrast to the momentum diffusion coefficients, the energy loss records an increasing trend with temperature. The sensitivity of the energy loss to temperature is greater for the Charm quark because of its smaller mass as compared to the bottom quark. The sensitivity to magnetic field is also greater for the Charm quark compared to the bottom quark, which is reflected in Fig.\eqref{dedxB01} by the discernible curves at $eB=0.05\, m_{\pi}^2$ and $eB=0.1\,m_{\pi}^2$. With regards to variation of HQ energy loss with magnetic field, the charm quark again records an opposite trend compared to the bottom quark, wherein the former decreases with increasing $B$, while the latter increases. Fig.\eqref{dedxB0} shows the variation of HQ energy loss with temperature in the absence of a background magnetic field. The temperature variation is the same as in the finite $B$ case, with both the magnitude as well as the rate of increase being greater for the charm quark. Again, this can be attributed to the lighter mass of the charm quark compared to bottom.
	
	\subsection{Case 2: $v\perp B$}
	\begin{figure}[H]
		\begin{subfigure}{0.48\textwidth}
			\includegraphics[width=0.96\textwidth]{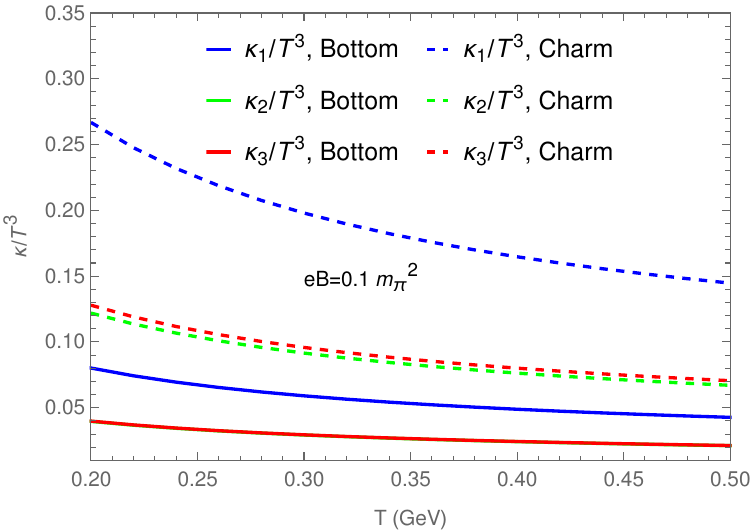}
			\caption{}\label{kperpB01}
		\end{subfigure}
		\hspace*{\fill}
		\begin{subfigure}{0.48\textwidth}
			\includegraphics[width=0.96\textwidth]{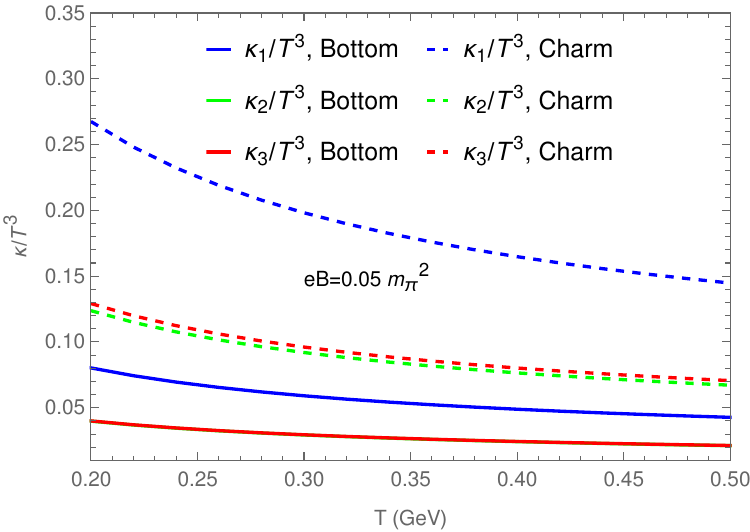}
			\caption{}\label{kperpB005}
		\end{subfigure}
		\begin{subfigure}{0.48\textwidth}
			\includegraphics[width=0.96\textwidth]{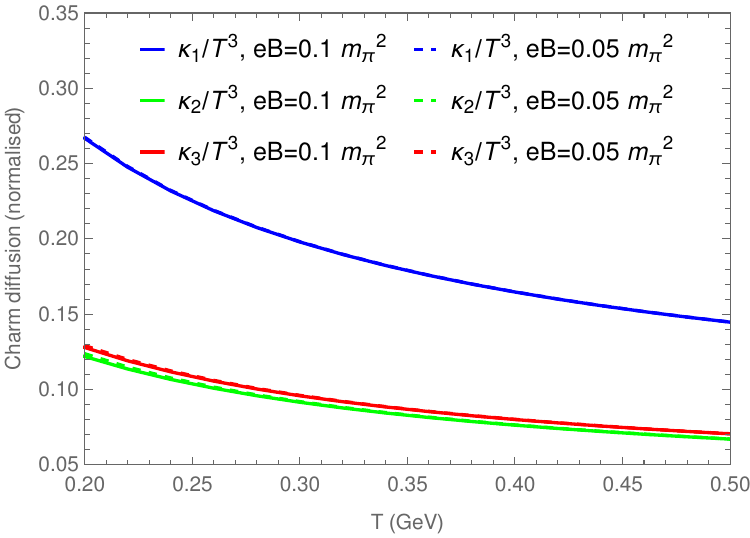}
			\caption{}\label{kperpcharm}
		\end{subfigure}
		\hspace*{\fill}
		\begin{subfigure}{0.48\textwidth}
			\includegraphics[width=0.96\textwidth]{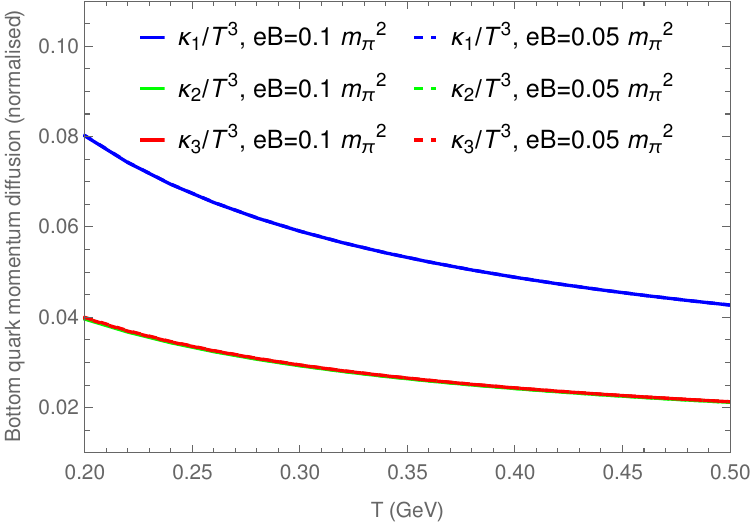}
			\caption{}\label{kperpbottom}
		\end{subfigure}
		\caption{(a) Temperature variation of normalised momentum diffusion coefficients of both Bottom and Charm quarks at $eB=0.1\,m_{\pi}^2$. (b) Same as (a) but at $eB=0.05\,m_{\pi}^2$. (c)Temperature variation of charm quark normalised momentum diffusion coefficients at different values of magnetic field strength. (d) Same as (c) but for Bottom quark. }
		\label{kperpQ}
	\end{figure}
 In the case of $\bm{v}\perp \bm{B}$, we have three momentum-diffusion coefficients $\kappa_{1}$, $\kappa_{2}$ and $\kappa_{3}$. In Figs. \eqref{kperpB01} and \eqref{kperpB005}, we show the variation of $\kappa_{1}$, $\kappa_{2}$ and $\kappa_{3}$ with $T$ at fixed values of the background magnetic field strength. For both the charm and bottom quarks, the $\kappa_1$ coefficient magnitude is significantly larger ($\sim$ 2 times) than $\kappa_2$ and $\kappa_3$. Also, like in the $\bm{v}\parallel\bm{B}$ case, the charm quark coefficients are larger in magnitude than their bottom quark counterparts. Overall, the momentum diffusion coefficients are smaller in magnitude than in the case of $\bm{v}\parallel\bm{B}$. Figs. \eqref{kperpcharm} and \eqref{kperpbottom} focus on the effect of magnetic field strength on the temperature variation of Charm and Bottom quark momentum diffusion coefficients, respectively. As is evident, the curves corresponding to different magnetic field strengths almost overlap. The variation with magnetic field strength is very feeble ($\sim 0.1\%$) for both the Charm and Bottom quarks. The momentum diffusion coefficients in this case are therefore much less sensitive to changes in magnetic field strength compared to the $\bm{v}\parallel\bm{B}$ case.

\begin{figure}[H]
	\begin{subfigure}{0.48\textwidth}
		\includegraphics[width=0.96\textwidth]{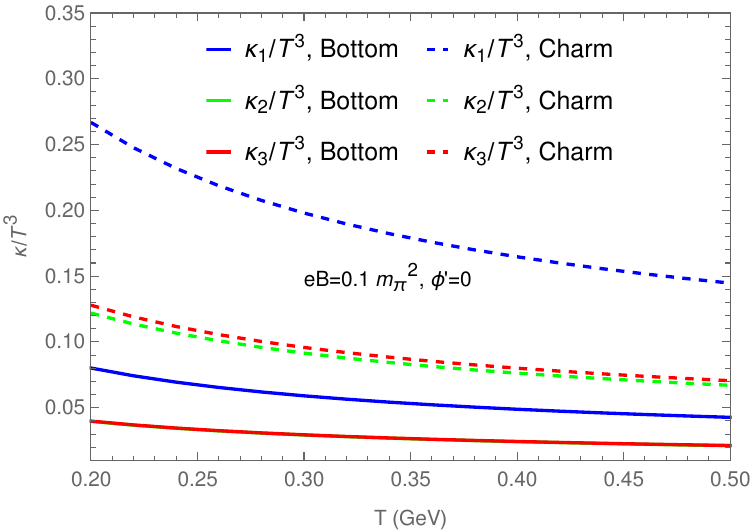}
		\caption{}\label{phi_0}
	\end{subfigure}
	\hspace*{\fill}
	\begin{subfigure}{0.48\textwidth}
		\includegraphics[width=0.96\textwidth]{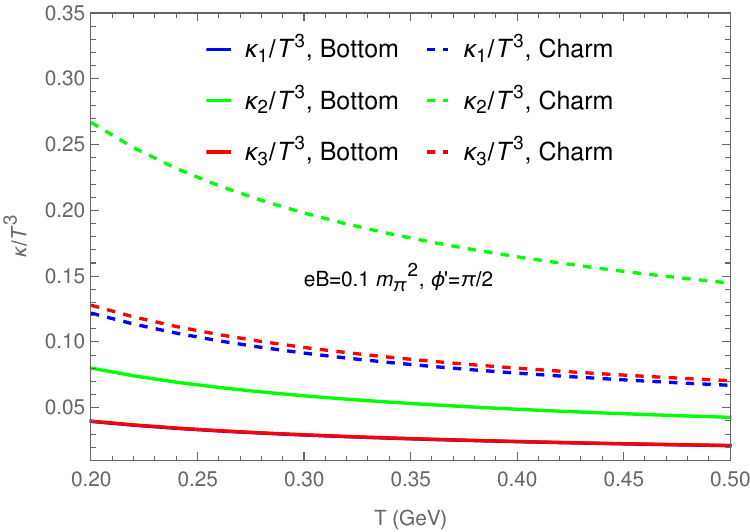}
		\caption{}\label{phi_2}
	\end{subfigure}
	\caption{$\phi'$ dependence of $\kappa$'s. (a) Momentum diffusion coefficients of both Charm and Bottom quarks at $\phi'=0$ and $eB=0.1\,m_{\pi}^2$ (b) Momentum diffusion coefficients of both Charm and Bottom quarks at $\phi'=\pi/2$ and $eB=0.1\,m_{\pi}^2$. }
	\label{figphi}
\end{figure}

\begin{figure}
	\centering
	\includegraphics[scale=0.65]{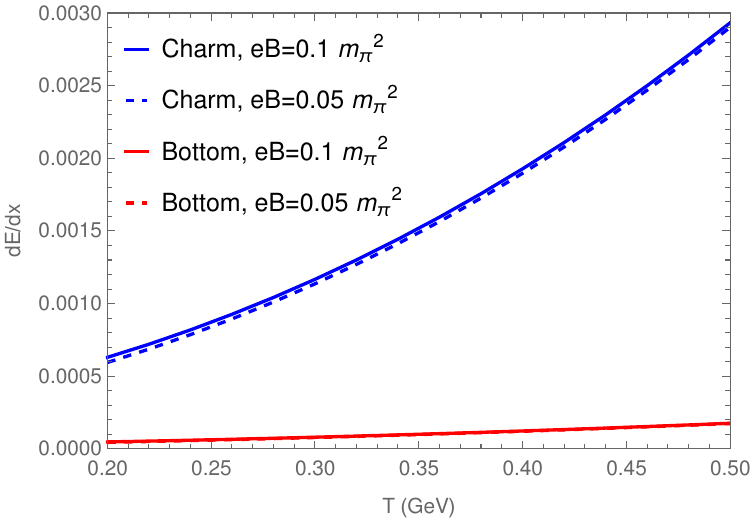}
	\caption{Energy loss of both Bottom and Charm quarks at different values of background magnetic field strengths.}
	\label{dedxperp}
\end{figure}

To specify the direction of HQ velocity, one needs to fix the value of $\phi'$.  Figs.[\ref{kperpQ}-\ref{dedxperp}] have been obtained with $\phi'=0$. From Eqs.(\ref{k1_perp}, \ref{k2_perp}), one can see that setting $\phi'=\pi/2$ leads to the following condition on the integrands\footnote{The limits of $y$ integration are $\phi'$ dependent, which also change as one changes $\phi'$. Thus, strictly, the equality in Eq.\eqref{phi_condition} holds only for the integrand, not for the integral. However, the value of the integral depends very weakly on the limits of the $y$ integration, so that, the final numerical values post integration satisfy Eq.\eqref{phi_condition} up to 5 significant figures after the decimal point.}
\begin{equation}
	\kappa_1(\phi'=\pi/2)=\kappa_2(\phi'=0)\,,\quad \kappa_2(\phi'=\pi/2)=\kappa_1(\phi'=0).\label{phi_condition}
\end{equation}
The condition \eqref{phi_condition} can be seen in Figs.(\ref{phi_0}, \ref{phi_2}), where, the blue and green curves are interchanged. It should be pointed out that this happens primarily because of the factors $\cos^2(y+\phi')$ and $\sin^2(y+\phi')$ in the $\kappa_{1}$ and $\kappa_{2}$ integrals [Eq.s \ref{k1_perp}, \ref{k2_perp}]. Physically, $\phi'=0$ and $\phi'=\pi/2$ correspond to the HQ velocity pointing towards $+\hat{x}$ and $+\hat{y}$, respectively. Also, $\kappa$ is a measure of mean squared momentum transfer between the HQ and the light thermal particles of the medium\cite{Moore:PRC71'2005}, so that $\kappa_{1/2/3}$ (as defined in Eqs.[\ref{k1}-\ref{k3}]) is a measure of the mean square of the $x/y/z$ component of the transfer momentum.  One can observe from Figs. \eqref{phi_0} and \eqref{phi_2} that when the HQ velocity points purely in the $+\hat{x}$ direction ($\phi'=0$), $\kappa_{1}$ is the largest. Similarly, when $\bm{v}$ points along $+\hat{y}$ ($\phi'=\pi/2$), $\kappa_{2}$ is the largest. This suggests that the momentum transfer between the HQ and the medium happens preferentially along the direction of HQ velocity.

Fig.\eqref{dedxperp} shows the variation of energy loss of both charm and bottom quarks with $T$, and also their dependence on magnetic field strength. Similar to the case 1, the energy loss is an increasing function of $T$, where, both the magnitude and the rate of increase being greater for the charm quark, compared to bottom quark. Compared to case 1, the sensitivity to changes in magnetic field strength is smaller; the charm quark values decrease by an average of 2.45\% as one goes from $eB=0.1\,m_{\pi}^2$ to $eB=0.05\,m_{\pi}^2$, while the corresponding value for Bottom quark is 2.74\%. Compared to the case of $\bm{v}\parallel\bm{B}$, the charm quark energy loss for $\bm{v}\perp\bm{B}$ is on an average smaller by 76.8\%.  The bottom quark values for case 2 is also smaller than those of case 1, on an average, by 94.3\%.

	\section{Application: Spatial diffusion coefficient}
	As mentioned in Section II, the drag coefficient $\eta$ can be obtained from the momentum diffusion coefficient via the fluctuation dissipation relation. 
	\begin{equation}
		\eta_D=\frac{\kappa}{2M_QT}.
	\end{equation}
	The zero momentum value of the drag coefficient is then obtained from the zero momentum value of the momentum diffusion coefficient
	\begin{equation}\label{drag_static}
		\eta_D(p=0)=\frac{\kappa(p=0)}{2M_QT}.
	\end{equation}
	The spatial diffusion coefficient $D_s$ can be defined via $\eta_D(p=0)$ as\cite{He:arxiv'2022}
	\begin{equation}\label{Ds}
		D_s=\frac{T}{\eta_D(p=0)M_q}.
	\end{equation}
	To evaluate $\kappa(p=0)$, we execute $p=0$ in the delta functions of Eqs.\eqref{kappa3L} and\eqref{kappa3T} which leads to $\omega=0$. So, we have to calculate the momentum diffusion coefficients in the $\omega\to 0$ limit.
	\begin{equation}
		\sum_{i=1}^4\frac{\rho_i(\omega,q)A_i(\omega,q)}{\omega}\bigg|_{\omega\to 0}=\frac{A_1\rho_1}{\omega}\bigg|_{\omega\to 0}
	\end{equation}
	All other terms vanish due to either the $A_i$'s or $\rho_i$'s vanishing in the $\omega\to 0$ limit. Finally, we are left with
	\begin{align}
		\kappa_L&=\frac{1}{2}\,\frac{\pi g^2T}{4M_Q^2(2\pi)^2}\int_0^{q_{\text{max}}} dq\int_0^\pi d\theta \frac{A_1(\omega=0)q^4}{\pi[q^4+q^2\,\Re_{b}(\omega=0)]^2}\,(q^4\sin^3\theta)\, \frac{\Im_{b}}{\omega}\bigg|_{\omega\to 0}\label{kappaLstatic}\\[0.3em]
		\kappa_T&=\frac{\pi g^2T}{4M_Q^2(2\pi)^2}\int_0^{q_{\text{max}}} dq\int_0^\pi d\theta \frac{A_1(\omega=0)q^4}{\pi[q^4+q^2\,\Re_{b}(\omega=0)]^2}\,(q^4\sin\theta \cos^2\theta)\, \frac{\Im_{b}}{\omega}\bigg|_{\omega\to 0}.\label{kappaTstatic}
	\end{align}
	The important thing to note is that $A_1(\omega=0)=M_Q^2$, and hence, the HQ mass dependence vanishes in $\kappa_L$ and $\kappa_{T}$. Thus, we expect the momentum diffusion coefficient values for the charm and bottom quarks to be identical.
	
	\begin{figure}[H]
		\centering
		\includegraphics[scale=0.7]{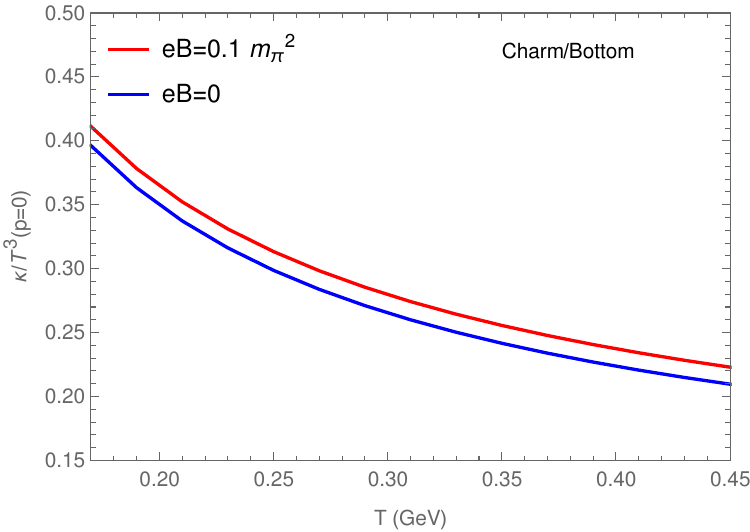}
		\caption{Normalised transverse and longitudinal momentum diffusion coefficients in the static limit ($\omega\to 0$) for both Charm and Bottom quarks as a function of temperature in the presence and absence of a background magnetic field.}
		\label{static}
	\end{figure}
	The momentum diffusion coefficient curves for the charm and bottom quarks overlap, as expected [Fig.\eqref{static}]. However, the longitudinal and transverse components are also indistinguishable. Hence,  when the heavy quarks are static, the momentum diffusion of HQ is isotropic, even in the presence of a background magnetic field.
	
	The case is similar for the spatial diffusion coefficient $D_s$ as well. From Eqs. \eqref{drag_static} and \eqref{Ds}, it can be seen that the HQ mass dependence cancels in $D_s$. In fact, this is one of the reasons why HQ diffusion is believed to carry generic information about the QCD medium.
	
	\vspace{5mm}
	\begin{figure}[H]
		\centering
		\includegraphics[scale=0.7]{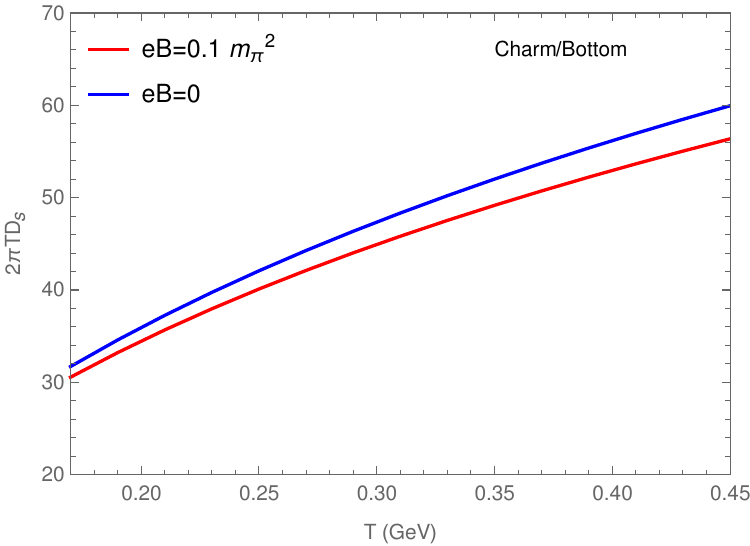}
		\caption{Spatial diffusion coefficient $D_s$ (multiplied with $2\pi T$) as a function of temperature in the presence and absence of a background weak magnetic field.}
		\label{Ds_fig}
	\end{figure}
	Fig.\eqref{Ds_fig} shows the variation of the scaled spatial diffusion coefficient (2$\pi T D_s$) with temperature in the presence of  background magnetic field of strength $eB=0.1m_{\pi}^2$. For comparison, the $B=0$ curve is also plotted. The curves corresponding to the charm and bottom quarks overlap, as expected. As can be seen, the spatial diffusion coefficient decreases in the presence of a magnetic field. This simply means that the mean squared momentum transfer per unit time between the HQ and the light partons (of which $\kappa$ is a measure) increases in the presence of a weak magnetic field, compared to $B=0$.
	The increasing trend with temperature is similar to what has been observed in several pQCD leading order (LO) studies in the past, both with $T$-dependent and $T$-independent couplings\cite{Svetitsky:PRD37'1988,Rapp:QGP4'2010,Moore:PRC71'2005}. 

	\section{Summary and Conclusions}
	In this work, we have investigated the dynamics of heavy quarks, \textit{viz.} Charm and Bottom in the presence of a weak background magnetic field. In particular, we have calculated the momentum diffusion coefficients and the energy loss perturbatively up to first order in the strong coupling $\alpha_s$. The interaction rate is calculated by considering $2\to 2$ elastic collisions of the form $Qq\to Qq$ and $Qg\to Qg$, by calculating the imaginary part of the heavy-quark self energy which is related to the squared matrix elements of the aforementioned collisional processes via the cutting rules. Gluon bremsstrahlung and Compton scattering processes are neglected since the former contributes only at higher order in $\alpha_s$ and the contribution of the latter is suppressed by powers of $M_Q/T$. There is a logarithmic U-V divergence present in the results of both the momentum diffusion coefficients and the energy loss. This is due to the fact that in this work, as a first attempt, we have calculated the contribution to the dynamical quantities arising out of only soft gluon exchange. It should however be remembered that the interaction rate is indeed dominated by processes involving soft gluon exchanges and hard scatterings contribute only at higher orders. In future, we shall include the hard scatterings too to get rid of the aforementioned U-V cut-off.
	
	 We have investigated the temperature dependence of the momentum diffusion coefficients and energy loss for both the heavy flavours, for two cases : 1) $\bm{v}\parallel\bm{B}$, and 2) $\bm{v}\perp\bm{B}$. In the former case, we define two momentum diffusion coefficients $\kappa_{L}$, $\kappa_T$, whereas in the latter, we have 3 such coefficients $\kappa_{1}$, $\kappa_{2}$, $\kappa_{3}$. The temperature behaviour of all the coefficients is similar in that the normalized coefficient magnitudes decrease with $T$.
	
	For case 1, $\kappa_{T}$ for both the flavours decreases with $T$ at the same rate, while $\kappa_L$ decreases faster for Charm. For the values of magnetic fields considered, the sensitivity of Charm quark diffusion coefficients to the magnetic field is found to be greater than that of the Bottom quark, possibly owing to the larger mass of the former.
	Further, the effect of increasing the magnetic field strength seems to have the opposite effects on the magnitudes of the Charm and Bottom diffusion coefficients; while the former decreases, the latter records an increase. For comparison, the $B=0$ results of the momentum diffusion coefficient ($\kappa$) have also been shown. It can be seen that the degree of anisotropy is much larger for the Charm quark than that of Bottom quark, which suggests that the mass of the heavy quark under consideration plays a strong role in determining the isotropicity of $\kappa$. For each of the flavors, $\kappa_L/T^3$ is found to be more sensitive to the magnetic field strength as compared to $\kappa_T/T^3$.  The heavy flavour energy loss is an increasing function of the temperature, both in the presence and absence of background magnetic field. Again, owing to its lighter mass, the sensitivity to both temperature and magnetic field is greater for the charmed quark, whereas for the Bottom quark, the curves corresponding to different magnetic fields almost overlap. The $B=0$ results are similar with the rate of increase of the charm quark energy loss being steeper.
	
	 For case 2, we observe certain similarities as well as differences vis-a-vis observations of case 1. For instance, similar to case 1, the rate of variation of $\kappa$ with $T$ is greater for Charm, compared to Bottom. Sensitivity of coefficient magnitudes to magnetic field strength is however much less for both the flavours in comparison with case 1. For both the flavours, $\kappa_{2}$ and $\kappa_{3}$ magnitudes, for $\phi'=0$, are almost similar for the entire temperature range (The similarity is more for Bottom), which in turn is significantly less than the $\kappa_{1}$ values. Thus, for $\phi'=0$, we have $|\kappa_{1}/T^3|>|\kappa_{2}/T^3|\approx|\kappa_{3}/T^3|$. Similarly, for $\phi'=\pi/2$, we obtain $|\kappa_{2}/T^3|>|\kappa_{1}/T^3|\approx|\kappa_{3}/T^3|$. Such a hierarchy in coefficient magnitudes, where, 2 of the coefficients are almost of the same magnitude, which in turn are significantly different from the third, is also seen in \cite{Bandyopadhyay:PRD105'2022}. The HQ energy loss exhibits a significant reduction in magnitude in comparison to case 1. The charm quark energy loss is on an average smaller by 76.8\%, whereas the bottom quark values for case 2 are smaller than those of case 1, on an average, by 94.3\%. Similar to case 1 however, the rate of increase of energy loss with $T$ is significantly larger for Charm quark, as compared to Bottom quark.
	
	We have also looked at the $p=0$ (static limit) results of momentum diffusion coefficients of the heavy quarks and found that the anisotropy in $\kappa$ completely vanishes in the said limit. Using this, we have looked at the spatial diffusion ($D_s$) of HQ. Interestingly, the HQ mass dependence cancels out both in $\kappa(p=0)$ and $D_s$. 
	
	\section{Acknowledgements}
	DD acknowledges the help of Aritra Bandyopadhyay received in course of fruitful discussions. DD also acknowledges the help received from Bithika Karmakar with regards to certain calculations pertaining to the evaluation of the resummed gluon propagator. DD is grateful to Abhishek Tiwary for checking several results presented here, and also for fruitful mathematical discussions relating to Sec V.B. DD also thanks Sumit for help in drawing Feynman diagrams.

	\appendix
	\appendixpage
	\addappheadtotoc
	\begin{appendices}
		\renewcommand \thesubsection{\Alph{section}.\arabic{subsection}}
		\section{Tensor structure of resummed gluon propagator in the presence of magnetic field}\label{AppendixA}
		We begin by discussing the 4-vectors that characterize the system under consideration. The fluid 4-velocity in local rest frame (LRF), and the metric tensor is given by
		\begin{equation}
			u^{\mu}=(1,0,0,0),\quad g^{\mu\nu}=\operatorname{diag}(1,-1,-1,-1).
		\end{equation}
		The direction of the external magnetic field is specified by the projection of the EM
		field tensor $F^{\mu\nu}$ along $u^\mu$:
		\begin{equation}
			n_{\mu}=\frac{1}{2B}\epsilon_{\mu\nu\rho\lambda}u^{\nu}F^{\rho\lambda}=(0,0,0,1)
		\end{equation}
		Introduction of these 4-vectors allows one to define a Lorentz invariant energy and momentum component as
		\begin{equation}
			q^0=q_0=Q\cdot u, \quad q^3=-q_3=Q\cdot n.
		\end{equation}
		We define parallel and perpendicular components of vectors and the metric tensor in the LRF as
		\begin{align}
			Q_{\parallel}^{\mu}&=(Q.u)u^{\mu}+(Q\cdot n)n^{\mu}=\left(q^0, 0,0, q^3\right);\\
			Q_{\perp}^{\mu}&=Q^{\mu}-Q_{\parallel}^{\mu}=\left(0, q^1,q^2, 0\right)\\
			Q_{\parallel}^2&=q_0^2-q_3^2\,, \quad Q_{\perp}^2=-(q_1^2+q_3^2)=-q_{\perp}^2\\[0.5em]
			g_{\|}^{\mu \nu}&=u^{\mu}u^{\nu}-n^{\mu}n^{\nu}=\operatorname{diag}(1,0,0-1), \\
			g_{\perp}^{\mu \nu}&=g^{\mu\nu}-g_{\|}^{\mu \nu}=\operatorname{diag}(0,-1,-1,0),
		\end{align}	
		One can further redefine $u^{\mu}$ and $n^{\mu}$ as
		\begin{align}
			\bar{u}^\mu&=u^\mu-\frac{(Q \cdot u) Q^\mu}{Q^2}=u^\mu-\frac{q_0 Q^\mu}{Q^2}\\
			\bar{n}^\mu&=n^\mu-\frac{(\tilde{Q} \cdot n) \tilde{Q}^\mu}{\tilde{Q}^2}=n^\mu-\frac{q_3 Q^\mu}{q^2}+\frac{q_0 q_3 u^\mu}{q^2},
		\end{align}	
		where, $\tilde{Q}^{\mu}=Q^{\mu}-(Q\cdot u)$. $\bar{u}^\mu$ and $\bar{n}^\mu$ so defined are orthogonal to $Q^{\mu}$ and $\tilde{Q}^{\mu}$, respectively. In the presence of a magnetic field, a set of basis tensors that are mutually orthogonal, can be constructed out of the 4-vectors mentioned above:
		\begin{align}
			\Delta_1^{\mu \nu}&=\frac{\bar{u}^\mu \bar{u}^\nu}{\bar{u}^2}, \label{d1}\\[0.3em]
			\Delta_2^{\mu \nu}&=g_{\perp}^{\mu \nu}-\frac{Q_{\perp}^\mu Q_{\perp}^\nu}{Q_{\perp}^2},\label{d2} \\[0.3em]
			\Delta_3^{\mu \nu}&=\frac{\bar{n}^\mu \bar{n}^\nu}{\bar{n}^2},\label{d3}\\[0.3em]
			\Delta_4^{\mu \nu}&=\frac{\bar{u}^\mu \bar{n}^\nu+\bar{u}^\nu \bar{n}^\mu}{\sqrt{\bar{u}^2} \sqrt{\bar{n}^2}}. \label{d4} 
		\end{align}
		These tensors satisfy the following properties:
		\begin{align}
			\left(\Delta_4\right)^{\mu \rho}\left(\Delta_4\right)_{\rho \nu}&=\left(\Delta_1\right)_\nu^\mu+\left(\Delta_3\right)_\nu^\mu, \\[0.3em]
			\left(\Delta_k\right)^{\mu \rho}\left(\Delta_4\right)_{\rho \nu}+\left(\Delta_4\right)^{\mu \rho}\left(\Delta_k\right)_{\rho \nu}&=\left(\Delta_4\right)_\nu^\mu, \\[0.3em]
			\left(\Delta_2\right)^{\mu \rho}\left(\Delta_4\right)_{\rho \nu}&=\left(\Delta_4\right)^{\mu \rho}\left(\Delta_2\right)_{\rho \nu}=0,
		\end{align}
		Any second rank tensor can be expanded in terms of these basis tensors. As such, the gluon self energy can be written as
		\begin{equation}\label{se}
			\Pi^{\mu\nu}(q_0,\bm{q})=b(q_0,\bm{q})\Delta_1^{\mu\nu}+c(q_0,\bm{q})\Delta_2^{\mu\nu}+d(q_0,\bm{q})\Delta_3^{\mu\nu}+a(q_0,\bm{q})\Delta_4^{\mu\nu},
		\end{equation}	
		where, $b$, $c$, $d$, $a$ are Lorentz invariant form factors. The Schwinger-Dyson equation relates the bare propagator, resummed propagator and the self energy of the particle under consideration. For the gluon propagator, we have
		\begin{equation}\label{sde}
			\mathcal{D}_{\mu\nu}^{-1}=\mathcal{D}_{\mu\nu}^0-\Pi_{\mu\nu},
		\end{equation}
		where, $\mathcal{D}_{\mu\nu}^0$ is the bare gluon propagator.
		We recall that any rank-2 tensor (and it's inverse) can be written in terms of the basis tensors $\Delta_i's$. Then, using Eq.\eqref{se}, \eqref{sde}, and the fact that 	$\mathcal{D}_{\mu\rho}^{-1}\mathcal{D}^{\rho\nu}=g_{\mu}^{\nu}$, we can derive the structure of the resummed gluon propagator as mentioned in Eq.\eqref{dmunu}

		\section{Form factors in weak magnetic field}\label{AppendixB}
		The fermion propagator in a weak background magnetic field is written as a series expansion in powers of $qB$ as (up to $\mathcal{O}(qB)^2$)
		\begin{align*}
			i S(K)&=  i \frac{\left(K+m_f\right)}{K^2-m_f^2}-q_f B \frac{\gamma_1 \gamma_2\left(K_{\|}+m_f\right)}{\left(K^2-m_f^2\right)^2} 
			-2 i\left(q_f B\right)^2 \frac{\left[K_{\perp}^2\left(K_{\|}+m_f\right)+K_{\perp}\left(m_f^2-K_{\|}^2\right)\right]}{\left(K^2-m_f^2\right)^4}\\
			&\equiv S_0(K)+S_1(K)+S_2(K)
		\end{align*}
		The quark loop (fermion) contribution to the gluon self energy is then given by
		\begin{align*}
			\Pi^{\mu \nu}_f(Q)= & -\sum_f \frac{i g^2}{2} \int \frac{d^4 K}{(2 \pi)^4} \operatorname{Tr}\left[\gamma^\nu\left\{S_0(K)+S_1(K)+S_2(K)\right\}\right.
			\left.\times \gamma^\mu\left\{S_0(P)+S_1(P)+S_2(P)\right\}\right]\\
			=&\Pi_{(0,0)}^{\mu \nu}(Q)+\Pi_{(1,1)}^{\mu \nu}(Q)+2 \Pi_{(2,0)}^{\mu \nu}(Q)+O\left[\left(q_f B\right)^3\right],
		\end{align*}
		where, 
		
		the first term is of $\mathcal{O}(qB)$ and the remaining are of $\mathcal{O}((qB)^2)$. The $\mathcal{O}(qB)$ term vanishes owing to Furry's theorem. The nonvanishing terms are given as:
		\begin{align}
			\Pi_{(0,0)}^{\mu \nu}(Q)&=\sum_f i 2 g^2 \int \frac{d^4 K}{(2 \pi)^4} \frac{\left[P^\mu K^\nu+K^\mu P^\nu-g^{\mu \nu}\left(K \cdot P-m_f^2\right)\right]}{\left(K^2-m_f^2\right)\left(P^2-m_f^2\right)}\label{se00}\\
			\Pi_{(1,1)}^{\mu \nu}(Q)&=\sum_f 2 i g^2\left(q_f B\right)^2 \int \frac{d^4 K}{(2 \pi)^4} \frac{\left[P_{\|}^\mu K_{\|}^\nu+K_{\|}^\mu P_{\|}^\nu+\left(g_{\|}^{\mu \nu}-g_{\perp}^{\mu \nu}\right)\left(m_f^2-K_{\|} \cdot P_{\|}\right)\right]}{\left(K^2-m_f^2\right)^2\left(P^2-m_f^2\right)^2}\label{se11}\\
			\Pi_{(2,0)}^{\mu \nu}(Q)&=-\sum_f 4 i g^2\left(q_f B\right)^2 \int \frac{d^4 K}{(2 \pi)^4}\left[\frac{M^{\mu \nu}}{\left(K^2-m_f^2\right)^4\left(P^2-m_f^2\right)}\right],\label{se20}
		\end{align}
		where, 
		\[M^{\mu\nu}=K_{\perp}^2\big[P^{\mu}K^{\nu}_{\parallel}+K^{\mu}_{\parallel}P^{\nu}-g^{\mu\nu}\left(K_{\parallel}\cdot P-m_f^2\right)\big]+(m_f^2-K_{\parallel}^2) \big[P^{\mu}K^{\nu}_{\perp}+K^{\mu}_{\perp}P^{\nu}-g^{\mu\nu}\left(K_{\perp}\cdot P\right)\big]\]
		
		The complete gluon self energy is expressed as:
		\begin{equation*}
			\Pi^{\mu \nu}(Q)=\Pi^{\mu \nu}_{\text{YM}}(Q)+\Pi^{\mu \nu}_f(Q),
		\end{equation*}
		where, $\Pi^{\mu \nu}_{\text{YM}}$ refers to the Yang-Mills contribution to the gluon self energy coming from the ghost and gluon loops, which is unaffected by the magnetic field. It is given by
		\begin{equation*}
			\Pi^{\mu \nu}_{\text{YM}}(Q)=-\frac{N_c\, g^2 T^2}{3} \int \frac{d \Omega}{2 \pi}\left(\frac{q_0 \hat{K}^\mu \hat{K}^\nu}{\hat{K} \cdot Q}-g^{\mu 0} g^{\nu 0}\right).
		\end{equation*}
		Using the properties of the tensors $\Delta_i$, the form factors can be expressed as:
		\begin{align}
			b(Q)&=\Delta_1^{\mu\nu}(Q)\Pi_{\mu\nu}(Q)=\Delta_1^{\mu\nu}(\Pi_{\mu\nu}^{\text{YM}}+\Pi_{\mu\nu}^f)=b_{\text{YM}}(Q)+b_{0f}(Q)+b_{2f}(Q)\\
			c(Q)&=\Delta_2^{\mu\nu}(Q)\Pi_{\mu\nu}(Q)=\Delta_2^{\mu\nu}(\Pi_{\mu\nu}^{\text{YM}}+\Pi_{\mu\nu}^f)=c_{\text{YM}}(Q)+c_{0f}(Q)+c_{2f}(Q)\\
			d(Q)&=\Delta_3^{\mu\nu}(Q)\Pi_{\mu\nu}(Q)=\Delta_3^{\mu\nu}(\Pi_{\mu\nu}^{\text{YM}}+\Pi_{\mu\nu}^f)=d_{\text{YM}}(Q)+d_{0f}(Q)+d_{2f}(Q)\\
			a(Q)&=\frac{1}{2}\Delta_4^{\mu\nu}(Q)\Pi_{\mu\nu}(Q)=\frac{1}{2}\Delta_4^{\mu\nu}(\Pi_{\mu\nu}^{\text{YM}}+\Pi_{\mu\nu}^f)=a_{\text{YM}}(Q)+a_{0f}(Q)+a_{2f}(Q).
		\end{align}
		In terms of powers of $qB$, the form factors can be expressed as
		\begin{equation}
			F(Q)=F_0(Q)+F_2(Q)=\left[F_{\text{YM}}(Q)+F_{0f}(Q)\right]+F_{2f}(Q), \quad F\equiv b,c,d,a
		\end{equation}
		
		\subsection{$\mathcal{O}(qB)^0$ terms of form factors}
		Using Eqs.(\ref{d1}-\ref{d4}) and Eq.\eqref{se}, we can write
		\begin{equation*}
			\Delta_1^{00}=\bar{u}^2, \quad \Delta_2^{00}=\Delta_3^{00}=\Delta_4^{00}=0, \quad \Pi^{00}=b\bar{u}^2
		\end{equation*}
		Thus, 
		\begin{equation}
			b_0(Q)=\frac{1}{\bar{u}^2}\left[\Pi^{\text{YM}}_{00}(Q)+\Pi^{(0,0)}_{00}(Q)\right]
		\end{equation}
		In the HTL approximation ($K\sim T$, $Q\sim gT$)
		\begin{equation}
			\Pi^{(0,0)}_{00}(Q)=\frac{N_f\,g^2T^2}{6}\left(1-\frac{q_0}{2q}\log\frac{q_0+q}{q_0-q}\right),\quad \Pi^{\text{YM}}_{00}=\frac{N_c\,g^2T^2}{3}\left(1-\frac{q_0}{2q}\log\frac{q_0+q}{q_0-q}\right)
		\end{equation}
		Thus, 
		\begin{equation}
			b_0(Q)=\frac{m_D^2}{\bar{u}^2}\left(1-\frac{q_0}{2q}\log\frac{q_0+q}{q_0-q}\right),
		\end{equation}
		where, 
		$m_D^2=\left.\left(\Pi_{00}^{(0,0)}+\Pi^{\text{YM}}_{00}\right)\right|_{\substack{p_0=0 \\ \mathbf{p} \rightarrow 0}}=\frac{g^2T^2}{3}\left(N_c+\frac{N_f}{2}\right)$ is the QCD Debye screening mass in the absence of magnetic field.
		
		An alternate way of evaluating the form factors is to calculate the self energy diagrammatically. As an example, the quark loop contribution to  $c_0$ will be evaluated this way.
		\begin{equation}
			c_0^f(Q)=\left(g_{\perp}^{\mu \nu}-\frac{Q_{\perp}^\mu Q_{\perp}^\nu}{Q_{\perp}^2}\right)\Pi^{(0,0)}_{\mu\nu}=T1-T2,
		\end{equation}
		where, 
		\begin{equation*}
			T1=g_{\perp}^{\mu \nu}\Pi^{(0,0)}_{\mu\nu}, \quad T_2=\frac{Q_{\perp}^\mu Q_{\perp}^\nu}{Q_{\perp}^2}\Pi^{(0,0)}_{\mu\nu}
		\end{equation*}
		Using the expression of $\Pi^{(0,0)}_{\mu\nu}$ from Eq.\eqref{se00} under the HTL approximation, we get,
		\begin{equation}
			T1=-\sum_{f}4 g^2(I_1-I_2),
		\end{equation}
		where,
		\begin{align}
			I_1=\int\frac{d^3k}{(2\pi)^3}T\sum_{n}\frac{K_{\perp}^2}{(K^2-m_f^2)(P^2-m_f^2)}\\
			I_2=\int\frac{d^3k}{(2\pi)^3}T\sum_{n}\frac{K^2}{(K^2-m_f^2)(P^2-m_f^2)}
		\end{align}
		$I_2$ is a well known integral which, under the HTL approximation, and in the limit of $m_f\to 0$ is $T^2/24$. The $I_1$ integral after summing over Matsubara frequencies simplifies to
		\begin{equation}
			I_1=-\frac{1}{2}\int\frac{d^3k}{(2\pi)^3}\left[\frac{n_F(E_1)}{E_1}-\left(1-\frac{q_0}{q_0-q\cos\theta}\right)\frac{dn_F(E_1)}{dk}\right](\cos^2\theta-1)\label{i_1}
		\end{equation}
		Here, $E_1\approx k$ is the energy of the fermion propagator having 4-momentum $K$, $\theta$ is the polar angle made by $\bm{k}$, and $n_F$ is the Fermi-Dirac distribution. First term of Eq.\eqref{i_1} evaluates to $T^2/72$. The second term can be expanded as
		\begin{equation*}
			\frac{1}{8\pi^2}\int dk\,d(\cos\theta)\,k^2 \frac{dn_F(k)}{dk}\left[\cos^2\theta-1-\frac{q_0\cos^2\theta}{q_0-q\cos\theta}+\frac{q_0}{q_0-q\cos\theta}\right].
		\end{equation*}
		This evaluates term by term to
		\[-T^2/72+T^2/24-\frac{q_0T^2}{48 q}\left[2\frac{q_0}{q}+\frac{q_0^2}{q^2}\log\frac{q_0+q}{q_0-q}-\log\frac{q_0+q}{q_0-q}\right]\].
		Thus, 
		\begin{equation}
			I_1-I_2=-\frac{T^2}{48 q^2}\left[2q_0^2-(q_0^2-q^2)\frac{q_0}{q}\log\frac{q_0+q}{q_0-q}\right]
		\end{equation}
		Similarly, it can be shown that 
		\begin{equation}
			T2=-\sum_{f}2 g^2(-I_1+I_2+I_3),
		\end{equation}
		where, 
		\[I_3(Q)=\int \frac{d^3k}{(2\pi)^3}T\sum_n\frac{2k_1k_2}{(K^2-m_f^2)(P^2-m_f^2)}=0\]
		Thus, 
		\begin{equation}
			c_0^f(Q)=T1+T2=-\sum_{f}2g^2[I_1-I-2]=\frac{N_f\,g^2T^2}{6}\frac{1}{2q^2}\left[q_0^2-(q_0^2-q^2)\frac{q_0}{2q}\log\frac{q_0+q}{q_0-q}\right]
		\end{equation}
		The Yang-Mills contribution is given by
		\[\frac{N_c\,g^2T^2}{3}\frac{1}{2q^2}\left[q_0^2-(q_0^2-q^2)\frac{q_0}{2q}\log\frac{q_0+q}{q_0-q}\right]\]
		Hence, finally, 
		\begin{equation}
			c_0(q_0,q)=\frac{m_d^2}{2q^2}\left[q_0^2-(q_0^2-q^2)\frac{q_0}{2q}\log\frac{q_0+q}{q_0-q}\right]
		\end{equation}
		Next, we have
		\begin{equation}
			d_0(q_0,q)=\frac{\bar{n}^\mu \bar{n}^\nu}{\bar{n}^2}\left(\Pi_{\mu\nu}^{\text{YM}}+\Pi_{\mu\nu}^{(0,0)}\right)
		\end{equation}
		It turns out that
		\begin{equation}
			d_0(q_0,q)=c_0(q_0,q)=\frac{m_d^2}{2q^2}\left[q_0^2-(q_0^2-q^2)\frac{q_0}{2q}\log\frac{q_0+q}{q_0-q}\right]
		\end{equation}
		The form factor $a_0$ is given by
		\begin{align}
			a_0(q_0,q)&=\frac{1}{2}\Delta_4^{\mu\nu}(\Pi_{\mu\nu}^{\text{YM}}+\Pi_{\mu\nu}^{(0,0)})\\\nonumber
			&=\frac{1}{2 \sqrt{\bar{u}^2} \sqrt{\bar{n}^2}}\left[-2 \frac{\bar{u} \cdot n}{\bar{u}^2}\left[\Pi_{00}^\text{YM}+\Pi_{00}^{(0,0)}\right]+2\left[\Pi_{03}^\text{YM}+\Pi_{03}^{(0,0)}\right]\right]\\\nonumber
			&=0
		\end{align}

		\subsection{$\mathcal{O}(qB)^2$ terms of form factors}
		\begin{equation}
			b_2(q_0,q)=\frac{u^{\mu}u^{\nu}}{\bar{u}^2}\left[\Pi_{\mu\nu}^{(1,1 )}+2\Pi_{\mu\nu}^{(2,0)}\right]
		\end{equation}
		Using Eqs. (\ref{se11}) and (\ref{se20}) in the above equation, we get
		\begin{equation*}
			b_2(q_0,q)=-\sum_f \frac{2 g^2\left(q_f B\right)^2}{\bar{u}^2} \int \frac{d^3 k}{(2 \pi)^3} T \sum_n\left\{\frac{\left.K^2+k^2\left(1+\cos ^2 \theta\right)+m_f^2\right)}{\left(K^2-m_f^2\right)^2\left(P^2-m_f^2\right)^2}+\frac{8\left(k^4+k^2 K^2\right)\left(1-\cos ^2 \theta\right)}{\left(K^2-m_f^2\right)^4\left(P^2-m_f^2\right)}\right\}
		\end{equation*}
		We make use of the HTL simplifications mentioned in Appendix C of \cite{Karmakar:EPJC79'2019} to further simplify $b_2$ to obtain
		\begin{align*}
			b_2=&-\sum_f \frac{2 g^2\left(q_f B\right)^2}{\bar{u}^2} \int \frac{d^3 k}{(2 \pi)^3} T \sum_n\Bigg\{\frac{1}{\left(K^2-m_f^2\right)^2\left(P^2-m_f^2\right)}+\frac{(-7+9c^2)k^2+2m_f^2}{\left(K^2-m_f^2\right)^3\left(P^2-m_f^2\right)}\\&-\frac{8(1-c^2)(k^4+m_f^2k^2)}{\left(K^2-m_f^2\right)^4\left(P^2-m_f^2\right)}\Bigg\},
		\end{align*}
		where, $c=\cos\theta$. Next, we perform the frequency sum using
		\begin{align*}
			T \sum_n \frac{1}{\left(\omega_n^2+E_k^2\right)\!\left[\left(\omega_n-\omega\right)^2+E_{k-q}^2\right]}&=  \frac{\left[1-n_F\left(E_k\right)-n_F\left(E_{k-q}\right)\right]}{4 E_k E_{k-q}}\!\left\{\frac{1}{i \omega+E_k+E_{k-q}}-\frac{1}{i \omega-E_k-E_{k-q}}\right\} \\
			& +\frac{\left[n_F\left(E_k\right)-n_F\left(E_{k-q}\right)\right]}{4 E_k E_{k-q}}\left\{\frac{1}{i \omega+E_k-E_{k-q}}-\frac{1}{i \omega-E_k+E_{k-q}}\right\},
		\end{align*}
		where, $E_k=\sqrt{k^2+m_f^2}$, $E_{k-q}=\sqrt{(k-q)^2+m_f^2}$. We write the expression in terms of mass derivatives to finally obtain
		\begin{align*}
			b_2\left(q_0, q\right)= & \sum_f \frac{2 g^2 q_f^2 B^2}{\bar{u}^2}\left\{\left(\frac{\partial^2}{\partial^2\left(m_f^2\right)}+\frac{5}{6} m_f^2 \frac{\partial^3}{\partial^3\left(m_f^2\right)}\right) \int \frac{d^3 k}{(2 \pi)^3} \frac{n_F\left(E_k\right)}{E_k}\left(\frac{q_0}{q_0-q \cos \theta}-1\right)\right. \\
			& +\left(\frac{\partial}{\partial\left(m_f^2\right)}+\frac{5}{6} m_f^2 \frac{\partial^2}{\partial^2\left(m_f^2\right)}\right) \int \frac{d^3 k}{(2 \pi)^3} \frac{n_F\left(E_k\right)}{2 E_k^3}\left(\frac{q_0}{q_0-q \cos \theta}\right) \\
			& -\left(\frac{\partial^2}{\partial^2\left(m_f^2\right)}+\frac{m_f^2}{2} \frac{\partial^3}{\partial^3\left(m_f^2\right)}\right) \int \frac{d^3 k}{(2 \pi)^3} \frac{n_F\left(E_k\right)}{E_k} \cos ^2 \theta\left(\frac{q_0}{q_0-q \cos \theta}-1\right) \\
			& \left.-\left(\frac{\partial}{\partial\left(m_f^2\right)}+\frac{m_f^2}{2} \frac{\partial^2}{\partial^2\left(m_f^2\right)}\right) \int \frac{d^3 k}{(2 \pi)^3} \frac{n_F\left(E_k\right)}{2 E_k^3} \cos ^2 \theta\left(\frac{q_0}{q_0-q \cos \theta}\right)\right\}
		\end{align*}
		After simplification, we finally obtain
		\begin{align}
			b_2= & \frac{\delta m_D^2}{\bar{u}^2}+\sum_f \frac{g^2\left(q_f B\right)^2}{\bar{u}^2 \pi^2}  \times\left[\left(g_k+\frac{\pi m_f-4 T}{32 m_f^2 T}\right)\left(A_0-A_2\right)\right.\nonumber \\
			& \left.+\left(f_k+\frac{8 T-\pi m_f}{128 m_f^2 T}\right)\left(\frac{5 A_0}{3}-A_2\right)\right] .
		\end{align}
		Here, $\delta m_D^2$ is the correction to the Debye mass due to weak magnetic field given by:
		\begin{align*}
			\delta m_D^2= &\left[\Pi_{\mu\nu}^{(1,1 )}+2\Pi_{\mu\nu}^{(2,0)}\right]_{q_0=0,\, q \rightarrow 0} =\sum_{f}\frac{g^2}{12\pi^2T^2}(q_fB)^2\sum_{l=1}^{\infty}(-1)^{l+1}l^2K_0\left(\frac{m_fl}{T}\right)\\
			f_k=&-\sum_{l=1}^{\infty}(-1)^{l+1} \frac{l^2}{16 T^2} K_2\left(\frac{m_f l}{T}\right)\\
			g_k=&\sum_{l=1}^{\infty}(-1)^{l+1} \frac{l}{4 m_f T} K_1\left(\frac{m_f l}{T}\right) .\\
			A_0=&\int \frac{d \Omega}{4 \pi} \frac{q_0 c^0}{Q \cdot \hat{K}}=\frac{q_0}{2 q} \log \left(\frac{q_0+q}{q_0-q}\right)\\
			A_2=&\int \frac{d \Omega}{4 \pi} \frac{q_0 c^2}{Q \cdot \hat{K}}=\frac{q_0^2}{2 q^2}\left(1-\frac{3q_3^2}{q^2}\right)\left(1-\frac{q_0}{2 q} \log \frac{q_0+q}{q_0-q}\right) +\frac{1}{2}\left(1-\frac{q_3^2}{q^2}\right) \frac{q_0}{2 q} \log \frac{q_0+q}{q_0-q}
		\end{align*}
		$K_0,$ $K_1$, $K_2$ are the modified Bessel functions of the second kind.
		Similarly, the form factor $c_2$ is given by
		\begin{equation}
			c_2(q_0,q)=\left(g_{\perp}^{\mu \nu}-\frac{Q_{\perp}^\mu Q_{\perp}^\nu}{Q_{\perp}^2}\right)\left[\Pi_{\mu\nu}^{(1,1 )}+2\Pi_{\mu\nu}^{(2,0)}\right]
		\end{equation}
		Using Eqs.\eqref{se11} and \eqref{se20}, we get
		\begin{align*}
			c_2(q_0,q)=&-
			\sum_f \frac{ g^2\left(q_f B\right)^2}{2} \int \frac{d^3 k}{(2 \pi)^3}T\sum_n 
			\left[\frac{4 k_0^2-4 k_3^2-4 m_f^2}{\left(K^2-m_f^2\right)^2\left(P^2-m_f^2\right)^2}+\frac{4\left(4 k_3^2-4 k_0^2+4 m_f^2\right)}{\left(K^2-m_f^2\right)^3\left(P^2-m_f^2\right)}\right. \\&
			\left.-\frac{4\left(k_0^2-k_3^2-m_f^2\right)\left(8 k_{\perp}^2-4 K^2+4 m_f^2+8(\boldsymbol{k} \cdot \boldsymbol{q})_{\perp}^2 / q_{\perp}^2\right.}{\left(K^2-m_f^2\right)^4\left(P^2-m_f^2\right)}\right]
		\end{align*}
		Using HTL approximations to simplify as earlier, we write $c_2$ in terms of mass derivatives as earlier
		\begin{align*}
			c_2= &- \sum_f 2  g^2\left(q_f B\right)^2 \int \frac{d^3 k}{(2 \pi)^3}T\sum_n \left[\frac{1}{2}+\frac{1}{4}\left(1-\cos ^2 \theta\right) \cos ^2 \phi+\frac{7}{4} \sin ^2 \theta\left(1+\cos ^2 \phi\right)\right. \\
			& \left.-\frac{5}{4} \sin ^4 \theta\left(1+\cos ^2 \phi\right)\right]  \times \frac{\partial}{\partial\left(m_f^2\right)} \frac{1}{\left(K^2-m_f^2\right)\left(P^2-m_f^2\right)}
		\end{align*}
		After performing the frequency sum followed by the integral, we finally obtain
		\begin{align}
			c_2(q_0,q)= & -\sum_f \frac{4 g^2\left(q_f B\right)^2}{3 \pi^2} g_k +\frac{g^2\left(q_f B\right)^2}{2 \pi^2}\left(g_k+\frac{\pi m_f-4 T}{32 m_f^2 T}\right) \times\left[-\frac{7}{3} \frac{q_0^2}{q_{\perp}^2}+\left(2+\frac{3}{2} \frac{q_0^2}{q_{\perp}^2}\right) A_0\right.\nonumber \\
			& +\left(\frac{3}{2}+\frac{5}{2} \frac{q_0^2}{q_{\perp}^2}+\frac{3}{2} \frac{q_3^2}{q_{\perp}^2}\right) A_2-\frac{3 q_0 q_3}{q_{\perp}^2} A_1 \left.-\frac{5}{2}\left(1-\frac{q_3^2}{q_{\perp}^2}\right) A_4-\frac{5 q_0 q_3}{q_{\perp}^2} A_3\right],
		\end{align}
		The remaining $A$ integrals are 
		\begin{align*}
			A_1=&-\frac{q_0 q_3}{q^2}\left[1-\frac{q_0}{2 q} \log \left(\frac{q_0+q}{q_0-q}\right)\right] \\
			A_3= & \frac{q_0}{2 q} \frac{q_3}{q}\left(1-\frac{5}{3} \frac{q_3^2}{q^2}\right) -\frac{3}{2} \frac{q_0}{q} \frac{q_3}{q}\left(1-\frac{q_0^2}{q^2}-\frac{q_3^2}{q^2}+\frac{5}{3} \frac{q_0^2}{q^2} \frac{q_3^2}{q^2}\right)  \times\left(1-\frac{q_0}{2 q} \log \frac{q_0+q}{q_0-q}\right)\\
			A_4= & \frac{3}{8}\left(1-\frac{q_3^2}{q^2}\right)^2-\frac{q_0^2}{8 q^2}\left(1-\frac{5 q_3^2}{q^2}\right)^2+\frac{5}{3} \frac{q_0^2}{q^2} \frac{q_3^4}{q^4}  -\frac{3}{8}\left\{\left(1-\frac{q_0^2}{q^2}\right)^2-\frac{2 q_3^2}{q^2}\left(1-\frac{3 q_0^2}{q^2}\right)^2\right. \\
			& \left.+\frac{q_3^4}{q^4}\left(1-\frac{5 q_0^2}{q^2}\right)^2+\frac{8 q_0^4}{q^4} \frac{q_3^2}{q^2}\left(1-\frac{5 q_3^2}{3 q^2}\right)\right\}  \times\left(1-\frac{q_0}{2 q} \log \frac{q_0+q}{q_0-q}\right) .
		\end{align*}
		It should be noted that the imaginary parts of the form factors come from the imaginary parts of the $A_i'$s. We write down the final expressions. The detailed derivation can be found in \cite{Karmakar:EPJC79'2019}
		
		\begin{align}
			d_2(q_0,q)&=\frac{\bar{n}^\mu \bar{n}^\nu}{\bar{n}^2}\left[\Pi_{\mu\nu}^{(1,1 )}+2\Pi_{\mu\nu}^{(2,0)}\right]\\
			&=F_1+F_2\nonumber,
		\end{align}
		where, 
		\begin{align*}
			F_1= & -\sum_f \frac{g^2\left(q_f B\right)^2 q^2}{\pi^2 q_{\perp}^2}  \times\left[g_k\left\{\frac{q_0^2 q_3^2}{3 q^4}+\frac{A_0}{4}-\left(\frac{3}{2}+\frac{q_0^2 q_3^2}{q^4}\right) A_2+\frac{5}{4} A_4\right\}\right. +\left(\frac{\pi}{32 m_f T}-\frac{1}{8 m_f^2}\right) \\
			&  \times\left\{\frac{A_0}{4}-\left(\frac{3}{2}+\frac{q_0^2 q_3^2}{p^4}\right) A_2+\frac{5}{4} A_4\right\} -f_k \frac{q_0^2 q_3^2}{q^4}\left(\frac{14}{3}-5 A_0+A_2\right) \left.+\frac{q_0^2 q_3^2}{q^4} \frac{8 T-\pi m_f}{128 T m_f^2}\left(5 A_0-A_2\right)\right],\\[0.5em]
			F_2= & -\sum_f \frac{g^2\left(q_f B\right)^2}{6 \pi^2 m_f T} \frac{q^0 q^3}{q_{\perp}^2} \frac{1}{1+\cosh \frac{m_f}{T}} \times\left(\frac{3 A_1}{2}-A_3\right) .
		\end{align*}
		Finally, 
		\begin{align}
			a_2(q_0,q)&=\frac{1}{2}\left(\frac{\bar{u}^\mu \bar{n}^\nu+\bar{u}^\nu \bar{n}^\mu}{\sqrt{\bar{u}^2} \sqrt{\bar{n}^2}}\right)\left[\Pi_{\mu\nu}^{(1,1 )}+2\Pi_{\mu\nu}^{(2,0)}\right]\\
			&=G_1+G_2\nonumber,
		\end{align}
		\begin{align*}
			G_1= & \sum_f \frac{4 g^2\left(q_f B\right)^2}{2 \pi^2 \sqrt{\bar{u}^2} \sqrt{\bar{n}^2}} \times\left[\frac{q_0 q_3}{q^2}\left\{\left(\frac{2}{3}-A_0+A_2\right) g_k+\left(\frac{4}{3}-\frac{5 A_0}{3}+A_2\right) f_k\right\}\right. \\
			& +\left\{\left(-A_0+A_2\right) \frac{\pi m_f-4 T}{32 T m_f^2}\right. \left.\left.-\frac{1}{6}\left(5 A_0-3 A_2\right) \frac{8 T-\pi m_f}{64 T m_f^2}\right\}\right] .
		\end{align*}
		\begin{align*}
			G_2
			= & \sum_f \frac{g^2\left(q_f B\right)^2}{\sqrt{\bar{u}^2} \sqrt{\bar{n}^2} 6 \pi^2 m_f T\left(1+\cosh \frac{m_f}{T}\right)}\times\left(-5 A_1+4 A_3\right) .
		\end{align*}

		\section{Calculation of spectral functions, $\rho_i$}\label{AppendixC}
		The cut part of the spectral functions are evaluated from the discontinuity in the pieces of the gluon propagators, which in turn is given by their imaginary parts analytically continued to real values of energy
		\begin{align*}
			\rho_1(\omega, q)= & -\frac{1}{\pi} \operatorname{Im}\left(\left.\chi_1\right|_{q_0=\omega+i \epsilon}\right) \\
			= & -\frac{1}{\pi} \operatorname{Im}\left(\left.\frac{\left(Q^2-d\right)}{\left(Q^2-b\right)\left(Q^2-d\right)-a^2}\right|_{q_0=\omega+i \epsilon}\right) \\
			= & -\frac{1}{\pi D}\left[\Im_{b}\left(\Im_{d}^2+\Re_{d}^2+Q^4-2 Q^2 \Re_{d}\right)\right.\left.+2 \Im_{a} \Re_{a}\left(Q^2-\Re_{d}\right)+\Im_{d}\left(\Re_{a}^2-\Im_{a}^2\right)\right] .
		\end{align*}
		Here $\Im$ and $\Re$ respectively depict the imaginary and real parts of the form factors.
		\begin{align*}
			\rho_2(\omega, q)&=-\frac{1}{\pi} \operatorname{Im}\left(\left.\chi_2\right|_{q_0=\omega+i \epsilon}\right) \\
			&=-\frac{1}{\pi} \operatorname{Im}\left(\left.\frac{1}{\left(Q^2-c\right)}\right|_{q_0=\omega+i \epsilon}\right) =-\frac{1}{\pi}\left[\frac{\Im_{c}}{\Im_{c}^2-\left(Q^2-\Re_{c}\right)^2}\right].
		\end{align*}
		
		\begin{align*}
			\rho_3(\omega, q)&=-\frac{1}{\pi} \operatorname{Im}\left(\left.\chi_3\right|_{q_0=\omega+i \epsilon}\right) \\
			&=-\frac{1}{\pi} \operatorname{Im}\left(\left.\frac{\left(Q^2-b\right)}{\left(Q^2-b\right)\left(Q^2-d\right)-a^2}\right|_{q_0=\omega+i \epsilon}\right) \\
			&=-\frac{1}{\pi D}\left[\Im_{d}\left(\Im_{b}^2+\Re_{b}^2+Q^4-2 Q^2 \Re_{b}\right)\right. 	\left.+2 \Im_{a} \Re_{a}\left(Q^2-\Re_{b}\right)+\Im_{b}\left(\Re_{a}^2-\Im_{a}^2\right)\right].
		\end{align*}
		\begin{align*}
			\rho_4(\omega, q)&=-\frac{1}{\pi} \operatorname{Im}\left(\left.\chi_4\right|_{q_0=\omega+i \epsilon}\right) \\
			=&-\frac{1}{\pi} \operatorname{Im}\left(\left.\frac{a}{\left(Q^2-b\right)\left(Q^2-d\right)-a^2}\right|_{q_0=\omega+i \epsilon}\right) \\
			=&-\frac{1}{\pi D}\left[\Im_{a}\left\{-\Im_{b} \Im_{d}+\Re_{b} \Re_{d}+\Re_{a}^2+\Im_{a}^2+Q^4-Q^2\left(\Re_{b}+\Re_{d}\right)\right\}\right. \\&\left.+\Re_{a}\left(Q^2\left(\Im_{b}+\Im_{d}\right) \Im_{d} \Re_{b}-\Im_{b} \Re_{d}\right)\right] \text {. } 
		\end{align*}
		Here the denominator $D$ is expressed as
		\begin{equation*}
			\begin{aligned}
				D= & {\left[\left(-\Im_{b} Q^2-\Im_{d} Q^2+\Im_{d} \Re_{b}+\Im_{b} \Re_{d}-2 \Im_{a} \Re_{a}\right)^2\right.}  \left.+\left(-\Im_{b} \Im_{d}+\Im_{a}^2+\left(Q^2-\Re_{b}\right)\left(Q^2-\Re_{d}\right)-\Re_{a}^2\right)^2\right]
			\end{aligned}
		\end{equation*}
	\end{appendices}


\begin{thebibliography}{9}	
		\bibitem{Arsene:NPA757'2005} I. Arsene, \textit{et al.}, BRAHMS Collaboration, \href{https://www.sciencedirect.com/science/article/pii/S0375947405002770?via%3Dihub}{ Nucl. Phys. A \textbf{757}, 1 (2005)}
		
		\bibitem{Adams:NPA757'2005} J. Adams, \textit{et al.}, STAR Collaboration, \href{https://www.sciencedirect.com/science/article/pii/S0375947405005294?via%3Dihub}{ Nucl. Phys. A \textbf{757}, 102 (2005)}	
		
		\bibitem{QGP3}P. F. Kolb and U. W. Heinz, \textit{Quark-Gluon Plasma}, Vol \textbf{3}, page 63, World Scientific (2004).
		
		\bibitem{Shuryak:NPA750'2005}E. V. Shuryak, \href{https://doi.org/10.1016/j.nuclphysa.2004.10.022}{Nucl. Phys. A \textbf{750}, 64 (2005) .}
		
		\bibitem{Huovinen:ARNPS56'2006}P. Huovinen and P. V. Ruuskanen, \href{https://doi.org/10.1146/annurev.nucl.54.070103.181236}{Ann. Rev. Nucl. Part. Sci. \textbf{56}, 163 (2006).}
		
		\bibitem{Hirano:JPG36'2009}T. Hirano, \href{https://dx.doi.org/10.1088/0954-3899/36/6/064031}{J. Phys. G \textbf{36}, 064031 (2009).}
		
		\bibitem{Kovtun:PRL95'2005}P. K. Kovtun, D. T. Son and A. O. Starinets, \href{https://journals.aps.org/prl/abstract/10.1103/PhysRevLett.94.111601}{Phys. Rev. Lett. 94, 111601 (2005)}.
		
		\bibitem{Frawley:PR462'2008}A. D. Frawley, T. Ullrich, and R. Vogt, \href{https://doi.org/10.1016/j.physrep.2008.04.002}{Phys. Rep. \textbf{462},  125–175 (2008).}
		
		\bibitem{Levai:PRC51'1995}P. L\'evai, B. M\"uller, and X. N. Wang, \href{https://doi.org/10.1103/PhysRevC.51.3326}{Phys. Rev. C \textbf{51}, 3326 (1995).}
		
		\bibitem{Das:PLB768'2019}S. K. Das, S. Plumari, S. Chatterjee, J. Alam, F. Scardina, and V. Greco, \href{http://dx.doi.org/10.1016/j.physletb.2017.02.046}{Phys. Lett. B, \textbf{768}, 134955 (2019).}
		
		\bibitem{Heinz:ACP739'2004}U. Heinz, \href{https://doi.org/10.1063/1.1843595}{AIP Conf. Proc. \textbf{739}, 163–180 (2004).}
		
		\bibitem{He:arxiv'2022}M. He, H. V. Hees and R. Rapp, \href{https://arxiv.org/abs/2204.09299}{arxiv:2204.09299v1 [hep-ph], 2022.}
		
		\bibitem{Rapp:QGP4'2010}R. Rapp and H. V. Hees, \textit{Quark Gluon Plasma}, Vol \textbf{4}, page 111, World Scientific (2010)
		
		\bibitem{Tuchin:AdvHEP'2013} K. Tuchin, \href{https://www.hindawi.com/journals/ahep/2013/490495/}{Adv.High Energy Phys. , 490495 (2013)}.
		
		\bibitem{Tuchin:PRC82'2010}K. Tuchin, \href{https://journals.aps.org/prc/abstract/10.1103/PhysRevC.82.034904}{Phys. Rev. C \textbf{82}, 034904 (2010)}.
		
		\bibitem{Tuchin:PRC83'2011}K. Tuchin, \href{https://journals.aps.org/prc/abstract/10.1103/PhysRevC.83.017901}{Phys. Rev. C \textbf{83}, 017901 (2011)}.
		
		\bibitem{Marty:PRC88'2013}R. Marty, E. Bratkovskaya, W. Cassing, J. Aichelin and H. Berrehrah, \href{https://journals.aps.org/prc/abstract/10.1103/PhysRevC.88.045204}{Phys. Rev. C \textbf{88}, 045204 (2013)}.
		
		\bibitem{Ding:PRD83'2011}H.-T. Ding, A. Francis, O. Kaczmarek, F. Karsch, E. Laermann, and W. Soeldner, \href{https://journals.aps.org/prd/abstract/10.1103/PhysRevD.83.034504}{Phys. Rev. D \textbf{83}, 034504  (2011)}.
		
		\bibitem{Gupta:PLB597'2004}S. Gupta, \href{https://www.sciencedirect.com/science/article/pii/S0370269304008913?via%3Dihub}{Phys. Lett. B \textbf{597}, 57–62  (2004)}.
		
		\bibitem{Amato:PRL111'2013}A. Amato, G. Aarts, C. Allton, P. Giudice, S. Hands, and J.-I. Skullerud, \href{https://journals.aps.org/prl/abstract/10.1103/PhysRevLett.111.172001}{Phys. Rev. Lett. \textbf{111}  no. 17, 172001 (2013)}. 
		
		\bibitem{Aarts:PRL99'2007}G. Aarts, C. Allton, J. Foley, S. Hands, and S. Kim, \href{https://journals.aps.org/prl/abstract/10.1103/PhysRevLett.99.022002}{Phys. Rev. Lett. \textbf{99}, 022002  (2007)}. 
		
		\bibitem{Puglisi:PRD90'2014}A. Puglisi, S. Plumari, and V. Greco, \href{https://journals.aps.org/prd/abstract/10.1103/PhysRevD.90.114009}{Phys. Rev. D \textbf{90} (2014) 114009}.
		
		\bibitem{Greif:PRD90'2014}M. Greif, I. Bouras, C. Greiner, and Z. Xu, \href{https://journals.aps.org/prd/abstract/10.1103/PhysRevD.90.094014}{Phys. Rev. D \textbf{90} (2014) no. 9, 094014}. 
		
		\bibitem{Hattori:PRD96'2017}K. Hattori, X. G. Huang, D. H. Rischke, and D. Satow, \href{https://doi.org/10.1103/PhysRevD.96.094009}{Phys. Rev. D \textbf{96}, 094009 (2017).}
		
		\bibitem{Gupta:PLB597'2004}S. Gupta, \href{https://doi.org/10.1016/j.physletb.2004.05.079}{Phys. Lett. B \textbf{597}, 57 (2004).}
		
		
		
		
		
		\bibitem{Kharzeev:NPA803'2008}D. E. Kharzeev, L. D. McLerran, and H. J. Warringa, \href{https://doi.org/10.1016/j.nuclphysa.2008.02.298}{Nucl.
			Phys. A\textbf{803}, 227 (2008).}
		
		\bibitem{Kharzeev:PRD83'2011}D. E. Kharzeev and H. U. Yee, \href{https://doi.org/10.1103/PhysRevD.83.085007}{Phys. Rev. D \textbf{83}, 085007
			(2011).}	
		
		\bibitem{Newman:JHEP01'2006}G. M. Newman, \href{https://dx.doi.org/10.1088/1126-6708/2006/01/158}{J. High Energy Phys. \textbf{01} (2006) 158.	}
		
		\bibitem{Burnier:PRL107'2011}Y. Burnier, D. E. Kharzeev, J. Liao, and H. U. Yee, \href{https://doi.org/10.1103/PhysRevLett.107.052303}{Phys.
			Rev. Lett. \textbf{107}, 052303 (2011).}	
		
		\bibitem{Gorbar:PRD83'2011}E. V. Gorbar, V. A. Miransky, and I. A. Shovkovy, \href{https://doi.org/10.1103/PhysRevD.83.085003}{Phys.
			Rev. D \textbf{83}, 085003 (2011).}
		
		
		\bibitem{Alexandre:PRD63'2001}J. Alexandre, K. Farakos, and G. Koutsoumbas, \href{https://doi.org/10.1103/PhysRevD.63.065015}{Phys. Rev.
			D \textbf{63}, 065015 (2001).}
		
		\bibitem{Gusynin:PRD56'1997}V. P. Gusynin and I. A. Shovkovy, \href{https://doi.org/10.1103/PhysRevD.56.5251}{Phys. Rev. D \textbf{56}, 5251
			(1997).}
		
		\bibitem{Lee:PRD55'1997}D. S. Lee, C. N. Leung, and Y. J. Ng, \href{https://doi.org/10.1103/PhysRevD.55.6504}{Phys. Rev. D \textbf{55}, 6504
			(1997).}	
		
		\bibitem{Bali:JHEP02'2012}G. S. Bali, F. Bruckmann, G. Endrodi, Z. Fodor, S. D. Katz,
		S. Krieg, A. Schafer, and K. K. Szabo, \href{https://doi.org/10.1007/JHEP02(2012)044}{J. High Energy Phys. \textbf{02} (2012) 044.}
		
		\bibitem{Farias:PRC90'2014}	R. L. S. Farias, K. P. Gomes, G. I. Krein, and M. B. Pinto,
		\href{https://doi.org/10.1103/PhysRevC.90.025203}{Phys. Rev. C \textbf{90}, 025203 (2014).}
		
		\bibitem{Farias:EPJA53'2017}	R. L. S. Farias, V. S. Timoteo, S. S. Avancini, M. B. Pinto,
		and G. Krein, \href{https://doi.org/10.1140/epja/i2017-12320-8}{Eur. Phys. J. A \textbf{53}, 101 (2017).}
		
		\bibitem{Mueller:PRD91'2015}N. Mueller and J. M. Pawlowski, \href{https://doi.org/10.1103/PhysRevD.91.116010}{Phys. Rev. D \textbf{91}, 116010}
		(2015).	
		
		\bibitem{Ayala:PRD90'2014}A. Ayala, M. Loewe, A. Z. Mizher, and R. Zamora, \href{https://doi.org/10.1103/PhysRevD.90.036001}{Phys.
			Rev. D \textbf{90}, 036001 (2014).}	
		
		\bibitem{Ayala:PRD91'2015}A. Ayala, M. Loewe, and R. Zamora, \href{https://doi.org/10.1103/PhysRevD.91.016002}{Phys. Rev. D \textbf{91},
			016002 (2015).}
		
		\bibitem{Ayala:PLB759'2016}A. Ayala, C. A. Dominguez, L. A. Hernandez, M. Loewe,
		and R. Zamora, \href{https://doi.org/10.1016/j.physletb.2016.05.058}{Phys. Lett. B \textbf{759}, 99 (2016).}
		
		\bibitem{Aarts:JHEP02'2015}G. Aarts, C. Allton, A. Amato, P. Giudice, S. Hands, and
		J.-I. Skullerud, \href{https://doi.org/10.1007/JHEP02(2015)186}{J. High Energy Phys. \textbf{02} (2015) 186.}
		
		\bibitem{Ding:PRD94'2016}H.-T. Ding, O. Kaczmarek, and F. Meyer, \href{https://doi.org/10.1103/PhysRevD.94.034504}{Phys. Rev. D \textbf{94},
			034504 (2016).}
		
		\bibitem{Rath:PRD100'2019}S. Rath and B. K. Patra, \href{https://doi.org/10.1103/PhysRevD.100.016009}{Phys. Rev. D \textbf{100}, 016009 (2019).}
		
		\bibitem{Khan:PRD104'2021}S. A. Khan and B. K. Patra, \href{https://doi.org/10.1103/PhysRevD.104.054024}{Phys. Rev. D \textbf{104}, 054024
			(2021).}
		
		\bibitem{Li:PRD97'2018}S. Li and H. Yee, \href{https://doi.org/10.1103/PhysRevD.97.056024}{Phys. Rev. D \textbf{97}, 056024 (2018).}	
		
		\bibitem{Panday:arxiv}P. Panday and B. K. Patra, \href{https://doi.org/10.48550/arXiv.2211.12303}{arXiv:2211.12303v1 [nucl-th] (2022). }
		
		\bibitem{Kurian:PRD103'2021}M. Kurian, \href{https://doi.org/10.1103/PhysRevD.103.054024}{Phys. Rev. D \textbf{103}, 054024 (2021).}	
		
		\bibitem{Khan:PRD107'2023}S. A. Khan and B. K. Patra, \href{https://doi.org/10.1103/PhysRevD.107.074034}{Phys. Rev. D \textbf{107}, 074034 (2023).}
		
		\bibitem{Dey:PRD104'2021}D. Dey and B. K. Patra, \href{https://doi.org/10.1103/PhysRevD.104.076021}{Phys. Rev. D \textbf{104}, 076021 (2021)}
		
		\bibitem{Hasan:PRD102'2020}M. Hasan and B. K. Patra, \href{https://doi.org/10.1103/PhysRevD.102.036020}{Phys. Rev. D \textbf{102}, 036020
			(2020).}
		
		\bibitem{Fukushima:PRD93'2016}K. Fukushima, K. Hattori, Ho-Ung Yee, and Y. Yin, \href{https://doi.org/10.1103/PhysRevD.93.074028}{Phys. Rev. D, \textbf{93}, 074028 (2016).} 
		
		\bibitem{Singh:JHEP68'2020}B. Singh, S. Mazumder and H. Mishra, \href{https://doi.org/10.1007/JHEP05(2020)068}{J. High Energy Phys., \textbf{68} (2020)}.
		
		\bibitem{Bandyopadhyay:PRD105'2022} A. Bandyopadhyay, J. Liao, and H. Xing, \href{https://doi.org/10.1103/PhysRevD.105.114049}{Phys. Rev. D, \textbf{105}, 114049 (2022).} 
		
		\bibitem{Jamal:arxiv}M. Y. Jamal, J. Prakash, I. Nilima and A. Bandyopadhyay, \href{https://doi.org/10.48550/arXiv.2304.09851}{arXiv:2304.09851 [hep-ph] (2023).}	
		
		\bibitem{Thoma:NPB351'1991}M. H. Thoma and M. Gyulassy, \href{https://doi.org/10.1016/S0550-3213(05)80031-8}{Nucl. Phys. B \textbf{351}, 491 (1991).}
				
		\bibitem{Braaten:PRD44'1991}E. Braaten, and M. H. Thoma, \href{https://doi.org/10.1103/PhysRevD.44.1298}{Phys. Rev. D \textbf{44}, 1298 (1991).}
		
		\bibitem{Moore:PRC71'2005}G. D. Moore and D. Teaney, \href{https://doi.org/10.1103/PhysRevC.71.064904}{Phys. Rev. C \textbf{71}, 064904 (2005).}
		
		\bibitem{Monteno:JPG38'2011} M. Monteno, W. M. Alberico, A. Beraudo, A. De Pace, A. Molinari, M. Nardi, and F. Prino, \href{https://dx.doi.org/10.1088/0954-3899/38/12/124144}{J. Phys. G \textbf{38}, 124144 (2011).}
		
		\bibitem{He:PLB735'2014}M. He, R. J. Fries, and R. Rapp, \href{https://doi.org/10.1016/j.physletb.2014.05.050}{Phys. Lett. B \textbf{735}, 445 (2014).}
		
    	\bibitem{Beraudo:EPJC75'2015}	A. Beraudo, A. De Pace, M. Monteno, M. Nardi, and
		F. Prino, \href{https://doi.org/10.1140/epjc/s10052-015-3336-6}{Eur. Phys. J. C \textbf{75}, 121 (2015).}
		
		\bibitem{Das:PLB747'2015}S. K. Das, F. Scardina, S. Plumari, and V. Greco, \href{https://doi.org/10.1016/j.physletb.2015.06.003}{Phys. Lett. B, \textbf{747}, 260-264 (2015).}
		
		\bibitem{Sadofyev:PRD93'2016} A. V. Sadofyev and Y. Yin, \href{https://doi.org/10.1103/PhysRevD.93.125026}{Phys. Rev. D \textbf{93}, 125026 (2016).}
		
		\bibitem{Akamatsu:PRC92'2015}Y. Akamatsu, \href{https://doi.org/10.1103/PhysRevC.92.044911}{Phys. Rev. C \textbf{92}, 044911 (2015).}
		
		\bibitem{Andronic:EPJC76'2016}A. Andronic \textit{et al.}, \href{https://doi.org/10.1140/epjc/s10052-015-3819-5}{Eur. Phys. J. C \textbf{76}, 107 (2016).}
		
		\bibitem{Kurian:PRD100'2019}M. Kurian, S. K. Das, and V. Chandra, \href{}{Phys. Rev. D \textbf{100}, 074003 (2019).}
		
		\bibitem{Singh:arxiv}B. Singh, M. Kurian, S. Mazumder, H. Mishra, V. Chandra,
		and S. K. Das, \href{https://doi.org/10.48550/arXiv.2004.11092}{arXiv:2004.11092 [hep-ph]}
		
		\bibitem{Banerjee:PRD85'2012}D. Banerjee, S. Datta, and R. Gavai, \href{https://doi.org/10.1103/PhysRevD.85.014510}{Phys. rev. D \textbf{85}, 014510 (2012)}.
		
		\bibitem{Bouttefeux:JHEP150'2020}A. Bouttefeux and M. Laine, \href{https://doi.org/10.1007/JHEP12(2020)150}{J. High Energy Phys. \textbf{150} (2020).}
		
		\bibitem{Banerjee:arxiv}D. Banerjee, S. Datta and M. Laine, \href{https://doi.org/10.48550/arXiv.2204.14075}{arXiv:2204.14075 [hep-lat] (2022).}
		
		\bibitem{Caron-Huot:JHEP04'2009}S. Caron-Huot, M. Laine, and G. D. Moore, \href{https://dx.doi.org/10.1088/1126-6708/2009/04/053}{J. High Energy Phys. \textbf{04}, 053 (2009).}
		
		\bibitem{Singh:PRD100'2019}B. Singh, A. Abhishek, S. K. Das,  and H. Mishra , \href{https://doi.org/10.1103/PhysRevD.100.114019}{Phys. Rev. D, \textbf{100}, 114019 (2019).}
		
		
		
		
		\bibitem{Chatterjee:PLB798'2019}S. Chatterjee  and P. Bożek, \href{https://doi.org/10.1016/j.physletb.2019.134955}{Phys. Lett. B, \textbf{798}, 260-264 (2017).}
		
		\bibitem{Liu:PRD105'2021}Jun-Hong Liu, S. K. Das, V. Greco, and M. Ruggieri, \href{https://doi.org/10.1103/PhysRevD.103.034029}{Phys. Rev. D, \textbf{103}, 034029 (2021).}
		
		\bibitem{EPJA54'2018}S. Mr\'owczy\'nski, \href{https://doi.org/10.1140/epja/i2018-12478-5}{Eur. Phys. J. A (2018) \textbf{54}, 43 (2018).}
		
		\bibitem{Rajagopal:JHEP10'2015}K. Rajagopal and A. V. Sadofyev, \href{https://doi.org/10.1007/JHEP10(2015)018}{J. High Energy Phys. \textbf{10} (2015) 018.}
		
		\bibitem{Casalderrey-Solana:PRD74'2006}J. Casalderrey-Solana  and D. Teaney, \href{https://doi.org/10.1103/PhysRevD.74.085012}{Phys. Rev. D, \textbf{74}, 085012 (2006).}
		
		\bibitem{Caron-Huot:PRL100'2008}S. Caron-Huot and G. D. Moore, \href{https://doi.org/10.1103/PhysRevLett.100.052301}{Phys. Rev. Lett. \textbf{100}, 052301 (2008),} \href{https://dx.doi.org/10.1088/1126-6708/2008/02/081}{J. High Energy Phys., \textbf{02}, 081 (2008).}
		
		\bibitem{Kumar:PRC105'2022}A. Kumar, M. Kurian, S. K. Das, and V. Chandra, \href{https://doi.org/10.1103/PhysRevC.105.054903}{Phys. Rev. C, \textbf{105}, 054903 (2022).}	
		
		\bibitem{Prakash:arxiv}J. Prakash, V. Chandra, and S. K. Das, \href{https://doi.org/10.48550/arXiv.2306.07966}{arXiv:2306.07966 [hep-ph] (2023.)}
		
		\bibitem{Mazumder:arxiv}S. Mazumder, V. Chandra, and S. K. Das, \href{https://doi.org/10.48550/arXiv.2211.06985}{arXiv:2211.06985 [hep-ph] (2023).}
		
		
		\bibitem{Weldon:PRD28'2007}H. A. Weldon, \href{https://doi.org/10.1103/PhysRevD.28.2007}{Phys. Rev. D \textbf{28}, 2007 (1983).}
		
		\bibitem{Landau:1969} L.D. Landau and E.M. Lifshitz, \textit{Statistical physics}, Addison-Wesley (1969).
		
		\bibitem{Reif:book}F. Reif, \textit{Fundamentals of statistical and thermal physics}, McGRAW-HILL BOOK COMPANY, 1965.
		
		\bibitem{Thoma:arxiv}M. H. Thoma, \href{https://doi.org/10.48550/arXiv.hep-ph/0010164}{arXiv:hep-ph/0010164}.
		
		\bibitem{Karmakar:EPJC79'2019}B. Karmakar, A. Bandyopadhyay, N. Haque, and M. G. Mustafa, \href{https://doi.org/10.1140/epjc/s10052-019-7154-0}{Eur. Phys. J. C, \textbf{79}, 658 (2019).}
		
		\bibitem{Pisarski:NPB309'1988}R. D. Pisarski, \href{https://doi.org/10.1016/0550-3213(88)90454-3}{Nucl. Phys. B \textbf{309}, 476 (1988).}
		
		\bibitem{Pisarski:NPA498'1989}R. D. Pisarski, \href{https://doi.org/10.1016/0375-9474(89)90620-9}{Nucl. Phys. A \textbf{498}, 423-428 (1989)}.
		
		\bibitem{Bazavov:PRD86'2012}A. Bazavov, N. Brambilla, X. Garcia i Tormo, P. Petreczky,
		J. Soto, and A. Vairo, \href{https://doi.org/10.1103/PhysRevD.86.114031}{Phys. Rev. D \textbf{86}, 114031 (2012).}
		
		\bibitem{Hattori:PRD95'2017}K. Hattori, S. Li, D. Satow, and H.-U. Yee, \href{https://doi.org/10.1103/PhysRevD.95.076008}{Phys. Rev. D \textbf{95}, 076008 (2017).}
		
		\bibitem{Beraudo:NPA831'2009}A. Beraudo, A. De Pace, W. M. Alberico, and A. Molinari,
		\href{https://doi.org/10.1016/j.nuclphysa.2009.09.002}{Nucl. Phys. A \textbf{831}, 59 (2009).}
		
		\bibitem{Svetitsky:PRD37'1988}B. Svetitsky, \href{https://doi.org/10.1103/PhysRevD.37.2484}{Phys. Rev. D \textbf{37}, 2484 (1988).}
		
		
		
		
		
	\end{thebibliography}
\end{document}